\documentclass[fleqn,usenatbib]{mnras}
\usepackage{newtxtext,newtxmath}
\usepackage[T1]{fontenc}
\usepackage{graphicx}	
\usepackage{amsmath}	
\hypersetup{linkcolor=blue}          
\usepackage{enumitem}

\usepackage[dvipsnames]{xcolor}

\newcommand{\aref}[1]{\hyperref[#1]{Appendix~\ref*{#1}}}

\def\equationautorefname~#1\null{equation~(#1)\null}

\defcitealias{Zhu2023a}{Paper I}
\defcitealias{Lu2023a}{Paper II}

\newcommand{\kms}{km s$^{-1}$}

\usepackage{scalerel,tikz}
\usetikzlibrary{svg.path}
\definecolor{orcidlogocol}{HTML}{A6CE39}
\tikzset{orcidlogo/.pic={
 \fill[orcidlogocol] svg{M256,128c0,70.7-57.3,128-128,128C57.3,256,0,198.7,0,128C0,57.3,57.3,0,128,0C198.7,0,256,57.3,256,128z};
 \fill[white] svg{M86.3,186.2H70.9V79.1h15.4v48.4V186.2z}
 svg{M108.9,79.1h41.6c39.6,0,57,28.3,57,53.6c0,27.5-21.5,53.6-56.8,53.6h-41.8V79.1z M124.3,172.4h24.5c34.9,0,42.9-26.5,42.9-39.7c0-21.5-13.7-39.7-43.7-39.7h-23.7V172.4z}
 svg{M88.7,56.8c0,5.5-4.5,10.1-10.1,10.1c-5.6,0-10.1-4.6-10.1-10.1c0-5.6,4.5-10.1,10.1-10.1C84.2,46.7,88.7,51.3,88.7,56.8z};
}}
\newcommand\orcidicon[1]{\href{https://orcid.org/#1}{\mbox{\scalerel*{
\begin{tikzpicture}[yscale=-1,transform shape]
\pic{orcidlogo};
\end{tikzpicture}
}{|}}}}

\title[Dynamics vs. stellar population relations in MaNGA]{MaNGA DynPop -- III. Stellar dynamics versus stellar population relations in 6000 early-type and spiral galaxies: Fundamental Plane, mass-to-light ratios, total density slopes, and dark matter fractions}

\author[K. Zhu et al.]{
Kai Zhu\orcidicon{0000-0002-2583-2669}$^{1,2,3}$\thanks{E-mail: \url{kaizhu@nao.cas.cn}},
Shengdong Lu\orcidicon{0000-0002-6726-9499}$^{4,5}$,
Michele Cappellari\orcidicon{0000-0002-1283-8420}$^{6}$,
Ran Li\orcidicon{0000-0003-3899-0612}$^{1,2,3}$\thanks{E-mail: \url{ranl@bao.ac.cn}},
Shude Mao\orcidicon{0000-0001-8317-2788}$^{4}$,\and
Liang Gao$^{1,2,3,5}$,
Junqiang Ge$^{1}$
\\
$^{1}$National Astronomical Observatories, Chinese Academy of Sciences, 20A Datun Road, Chaoyang District, Beijing 100101, China\\
$^{2}$Institute for Frontiers in Astronomy and Astrophysics, Beijing Normal University, Beijing 102206, China\\
$^{3}$School of Astronomy and Space Science, University of Chinese Academy of Sciences, Beijing 100049, China\\
$^{4}$Department of Astronomy, Tsinghua University, Beijing 100084, China\\
$^{5}$Institute for Computational Cosmology, Department of Physics, University of Durham, South Road, Durham, DH1 3LE, UK\\
$^{6}$Sub-department of Astrophysics, Department of Physics, University of Oxford, Denys Wilkinson Building, Keble Road, Oxford, OX1 3RH, UK\\
}

\date{Accepted: 2023 October 17; Revised: 2023 October 16; Received: 2023 March 4}

\pubyear{2023}

\begin{document}
\label{firstpage}
\pagerange{\pageref{firstpage}--\pageref{lastpage}}
\maketitle

\begin{abstract}
We present dynamical scaling relations, combined with the stellar population properties, for a subsample of about 6000 nearby galaxies with the most reliable dynamical models extracted from the full Mapping Nearby Galaxies at Apache Point Observatory (MaNGA) sample of 10 000 galaxies. We show that the inclination-corrected mass plane for both early-type galaxies (ETGs) and late-type galaxies (LTGs), which links dynamical mass, projected half-light radius $R_{\rm e}$, and the second stellar velocity moment $\sigma_{\rm e}$ within $R_{\rm e}$, satisfies the virial theorem and is even tighter than the uncorrected one. We find a clear parabolic relation between $\lg(M/L)_{\rm e}$, the total mass-to-light ratio within a sphere of radius $R_{\rm e}$, and $\lg\sigma_{\rm e}$, with the $M/L$ increasing with $\sigma_{\rm e}$ and for older stellar populations. However, the relation for ETGs is linear and the one for the youngest galaxies is constant. We confirm and improve the relation between mass-weighted total density slopes $\overline{\gamma_{_{\rm T}}}$ and $\sigma_{\rm e}$: $\overline{\gamma_{_{\rm T}}}$ become steeper with increasing $\sigma_{\rm e}$ until $\lg(\sigma_{\rm e}/{\rm km\,s^{-1}})\approx 2.2$ and then remain constant around $\overline{\gamma_{_{\rm T}}}\approx 2.2$. The $\overline{\gamma_{_{\rm T}}}-\sigma_{\rm e}$ variation is larger for LTGs than ETGs. At fixed $\sigma_{\rm e}$ the total density profiles steepen with galaxy age and for ETGs. We find generally low dark matter fractions, median $f_{\rm DM}(<R_{\rm e})=8$ per cent, within a sphere of radius $R_{\rm e}$. However, we find that $f_{\rm DM}(<R_{\rm e})$ depends on $\sigma_{\rm e}$ better than stellar mass: dark matter increases to a median $f_{\rm DM}(<R_{\rm e})=33$ percent for galaxies with $\sigma_{\rm e}\la100$ \kms. The increased $f_{\rm DM}(<R_{\rm e})$ at low $\sigma_{\rm e}$ explains the parabolic $\lg(M/L)_{\rm e}-\lg\sigma_{\rm e}$ relation.
\end{abstract}




\section{Introduction}

The dynamical scaling relations connect the observables of galaxies, e.g. their mass or luminosity, their size, and their internal kinematics, providing key tests for galaxy formation theory. The most widely used dynamical scaling relations include the Tully-Fisher relation \citep[TF;][]{Tully1977} for the late-type galaxies (LTGs), the Faber-Jackson relation \citep[FJ;][]{Faber1976} for the early-type galaxies (ETGs; including ellipticals and lenticulars), and the Fundamental Plane \citep[FP;][]{Djorgovski1987,Dressler1987} which is extended from the FJ by including the galaxy size as a third parameter. The dynamical relations were proposed as distance estimators originally, but they also contain useful information about galaxy evolution. According to the hierarchical galaxy formation model, galaxies increase their mass and size through various processes (e.g. gas accretion induced star formation, and mergers with other galaxies), thus leaving imprint on the final observed relations. Therefore, the dynamical scaling relations provide strong constraints on the galaxy formation and evolution theory.

The initial papers on the FP and FJ focused on elliptical galaxies, while those on the TF were applied to spiral galaxies only. Later studies extended the samples to other galaxy morphological types. Some studies found that the dynamical scaling relations can be generalized to all ETGs \citep{Jorgensen1996,Cappellari2006,Cappellari2013a}. Moreover, when using different kinematic tracers for ETGs/LTGs, specifically stellar kinematics for ETGs and gas kinematics for LTGs, various authors proposed unified dynamics scaling relations valid for all galaxies \citep{Burstein1997,Zaritsky2008,Dutton2011,Cortese2014}. These kinds of generalized FP including both LTGs and ETGs were subsequently shown to hold, with even higher accuracy, when consistently using the same stellar kinematics tracer from IFS for all morphological types \citep{Lihongyu2018,Aquino-Ortiz2020,Ferrero2021}. The tight FP, which consists of luminosity, size, and velocity dispersion, was interpreted as due to the virial equilibrium, as originally suggested \citep{Faber1987}. But the reason for the deviation between the coefficients of the FP and the virial ones (known as the `tilt' of the FP) remained a source of debate for some time \citep[e.g.][]{Ciotti1996,Scodeggio1998,Pahre1998,Bernardi2003,Trujillo2004}.

With the advent of integral field spectroscopy (IFS) galaxy survey, e.g. SAURON \citep{deZeeuw2002}, $\rm ATLAS^{3D}$ \citep{Cappellari2011}, CALIFA \citep{Sanchez2012}, SAMI \citep{Bryant2015}, and MaNGA \citep{Bundy2015}, one can construct detailed dynamical models using the spatially resolved stellar kinematics and obtain accurate dynamical mass (or total mass-to-light ratio $M/L$) measurements. Using the stellar dynamical models, \citet{Cappellari2006} analysed 25 ETGs in the SAURON survey and found that the tilt of the FP is almost exclusively due to the variation of total $M/L$, while \citet{Cappellari2013a} confirmed this by replacing the luminosity of the FP with dynamical mass and obtaining the very tight Mass Plane (MP) for 260 $\rm ATLAS^{3D}$ ETGs. Independent confirmations by gravitational lensing \citep{Bolton2008,Auger2010} also support that the variation of total $M/L$ causes the tilt of the FP. More recently, the nearby IFS surveys with a large sample containing various types of galaxies (e.g. the SAMI survey and the MaNGA survey), as well as the higher-redshift ($z\sim0.8$) LEGA-C survey \citep{vanderWel2016}, also found the very tight MP satisfying the virial theorem for both the ETGs and LTGs \citep{Lihongyu2018,deGraaff2021,DEugenio2021}. This confirms the common origin of the MP for both the ETGs and LTGs, which is also consistent with the results found in cosmological simulations \citep{deGraaff2023}. However, given that the total $M/L$ is associated with the stellar mass-to-light ratio and the dark matter fraction, it is still worthy to investigate the separate contributions of the two sources, as well as the contribution of non-homology in light profiles \citep{Ciotti1996,Graham1997,Prugniel1997,Bertin2002,Trujillo2004,Bernardi2020}, to the tilt and the scatter of the FP. 

The completion of the MaNGA survey \citep{SDSSDR17}, which is the largest sample of galaxies ever observed with IFS and consists of data with radial coverage carefully matched to the galaxy sizes, motivates us to revisit the study of the FP, the MP, and the total $M/L$ of the MaNGA galaxies, using the quantities derived from the well-established Jeans Anisotropic Modelling (JAM) models (\citealt{Zhu2023a}, hereafter \citetalias{Zhu2023a}) and the Stellar Population Synthesis (SPS) models (\citealt{Lu2023a}, hereafter \citetalias{Lu2023a}).

In addition to the amount of total mass, other important quantities derived from the dynamical models are the total mass-density slope and the dark matter fraction. According to the current paradigm of hierarchical galaxy formation \citep{White1978}, galaxies are embedded in dark matter halos which can not be observed directly but still play an important role in the formation and evolution of galaxies. The dark matter fraction, which is usually defined as the ratio between the amount of dark matter mass and the total mass within an effective radius, gives a direct measurement of dark matter content but also suffers from strong degeneracy (see the discussion in sec.~6.3 of \citetalias{Zhu2023a}). We expect that the dark matter fractions are statistically correct but they should be used with caution. The total mass distribution, which combines the observed stellar mass distribution and the dark matter mass distribution, provides a robust quantity to understand the interplay between the two components. The scaling relations of total density slopes had been established from various methods and different samples: e.g. the stellar dynamics for the ETGs at large ($\sim4 R_{\rm e}$ half-light radii) \citep{Cappellari2015,Bellstedt2018} or small radii ($\sim1 R_{\rm e}$) \citep{Poci2017}, HI gas rotation curves in ETGs at large radii \citep{Serra2016}, gravitational lensing for the ETGs at small radii \citep{Bolton2008,Auger2010}, gas rotation curves for the LTGs \citep{Tortora2019}. Recently, \citet{Liran2019} found a unified relation between the total density slopes and the $\sigma_{\rm e}$ (or $M_{\ast}$) for both the ETGs and LTGs in MaNGA, while, however, a scatter of the total density slopes at fixed $\sigma_{\rm e}$ is presented especially at the low-$\sigma_{\rm e}$ end. The scaling relation of the total density slopes is worthy of further study using a larger sample and combining it with the stellar population properties. 

As discussed in the introduction of \citetalias{Zhu2023a}, both JAM and Schwarzschild methods show no systematic biases in recovering the total mass distribution, suggested by the detailed comparison between the two methods using observed \citep{Leung2018} and simulated galaxies \citep{Jin2019}. However, as opposed to previous thinking that more general dynamical models imply better accuracy, a smaller scatter for JAM is found in this case. When comparing with the observed CO gas circular velocities within a radial range of $0.8-1.6 R_{\rm e}$, where the gas kinematics is well resolved and the circular velocities are more robustly determined, the mean ratio for 54 galaxies between the errors of the Schwarzschild and of the JAM models is $\langle\sigma_{\rm SCH}/\sigma_{\rm JAM}\rangle\approx1.7$ \citep[fig.~8 and tab.~4]{Leung2018}. Similarly, when considering 45 model fits to the galaxies in numerical simulations, the 68th percentile (1$\sigma$ error) of absolute deviations between the recovered and the true enclosed masses (inside a sphere of $R_{\rm e}$) is a factor of 1.6 smaller for JAM than for Schwarazschild \citep[][fig.~4]{Jin2019}.

\citet{Quenneville2022} pointed out a small bug in the triaxial Schwarzschild code by \citet{vandenBosch2008} which was used in the two above studies. This could potentially affect the accuracy of the Schwarzschild results and explain the larger uncertainties than JAM. However, a response by \citet{Thater2022} concluded that any effect on previous results using that Schwarzschild code was insignificant.

Recently, \citet{Neureiter2023} found that one can improve the accuracy of Schwarzschild models using a simple data-driven optimization method developed in \citet{Thomas2022}. However, this result was only tested on a single simulation of a triaxial slow rotator, and has yet to be confirmed by independent groups. It is unclear whether it can be extended to the general class of fast rotators, which dominate the local Universe and the MaNGA sample. An independent analysis of a larger sample of real galaxies or simulations, directly comparing different methods in the very same conditions, as done by \citet{Leung2018} and \citet{Jin2019}, would be very valuable using the \textsc{smart} code used by \citet{Neureiter2021,Neureiter2023}. However, unlike the \citet{vandenBosch2008} code, upgraded and renamed \textsc{dynamite} by \citet{Thater2022}, the \textsc{smart} code has not yet been publicly released.

In this paper, we make use of the largest sample of IFS observations from the MaNGA survey, which includes various types of nearby galaxies, to study the dynamical scaling relations. The scaling relations use the quantities derived from the accurate JAM models \citepalias{Zhu2023a} and the stellar population properties \citepalias{Lu2023a}. Given the large sample, the various types of galaxies, and the well-established dynamical models with quality validation, we propose the relations presented in this paper as the benchmark for the dynamical scaling relations of nearby galaxies. The combination of dynamical scaling relations and stellar population properties also provides a novel view on galaxy formation and evolution.

The paper is organised as follows. In \autoref{sec:data_method}, we briefly introduce the MaNGA kinematic data, the JAM models, the SPS models, and the quantities we used in this work. We present the main results in \autoref{sec:results}, including the FP, the MP, the total M/L, the total density slopes, the dark matter fractions, and the dynamical properties on the mass-size plane. We present the discussions in \autoref{sec:discuss}. Finally, we summarize the results in \autoref{sec:summary}. Throughout the paper, we assume a flat Universe with $\Omega_{\rm m} = 0.307$ and $H_0 = 67.7\,\mathrm{km\,s^{-1}\,Mpc^{-1}}$ \citep{Planck2016}.

\section{Data and methods}
\label{sec:data_method}
\subsection{The MaNGA data and galaxy sample}
The Mapping Nearby Galaxies at Apache Point Observatory (MaNGA) survey \citep{Bundy2015} is one of three projects in Sloan Digital Sky Survey-IV \citep[SDSS-IV;][]{Blanton2017}, which provides spatially resolved spectral measurements for $\sim 10 000$ nearby galaxies. Using the integral field unit (IFU) technique, the MaNGA project simultaneously obtains the spectra across the face of target galaxies with the tightly-packed fiber bundles that feed into the BOSS spectrographs \citep{Smee2013,Drory2015} on the Sloan 2.5m telescope \citep{Gunn2006}. The field-of-view (FoV) of MaNGA observations covers a radial range out to 1.5 effective radii ($R_{\rm e}$) for $\sim 2/3$ galaxies (Primary+ sample) and out to 2.5 $R_{\rm e}$ for $\sim 1/3$ galaxies (Secondary sample) at higher redshift \citep{Law2015,Wake2017}.

The spectra of MaNGA span a wavelength range of $3600-10300\,\Angstrom$, with a spectral resolution of $\sigma = 72\, \rm {\rm km\,s^{-1}}$ \citep{Law2016}. Data cubes are produced by spectrophotometrically calibrating \citep{Yan2016} the raw data and then processing the calibrated data with the Data Reduction Pipeline \citep[DRP;][]{Law2016}. Stellar kinematic maps are extracted from the data cubes using the Data Analysis Pipeline \citep[DAP;][]{Belfiore2019,Westfall2019}, which uses the \textsc{ppxf} software \citep{Cappellari2004,Cappellari2017} with a subset of MILES stellar library \citep{Sanchez-Blazquez2006,Falcon-Barroso2011}, MILES-HC, to fit the absorption lines of IFU spectra. Before extracting stellar kinematics, the spectra are Voronoi binned \citep{Cappellari2003} to signal-to-noise ratio $\rm (S/N)=10$ to obtain reliable stellar velocity dispersions.

\subsection{Sample selection}

In the final data release of MaNGA \citep[SDSS DR17;][]{SDSSDR17}, there are 10296 galaxies if the targets of ancillary programs (the Coma, IC342, M31, and globular clusters) are excluded from the total 10735 DAP outputs. We derive the dynamical properties for 10296 galaxies using the Jeans Anisotropic Modelling \citep[JAM;][]{Cappellari2008,Cappellari2020} in \citetalias{Zhu2023a}. The whole sample is classified as different modelling qualities (Qual = $-1$, 0, 1, 2, 3 from worst to best) based on the comparisons between observed and modelled stellar kinematics. In this work, we select 6065 galaxies that are flagged as $\rm Qual \geqslant 1$, for which the dynamical quantities related to the total mass distribution are nearly insensitive to different model assumptions \citepalias{Zhu2023a}. The adopted subset of models are those for which we estimated that both zeroth-order quantities like the total mass and $M/L$ and first-order quantities like the total density slope can be trusted. In sec.~5.1 of \citetalias{Zhu2023a}, we explained why we excluded the galaxies with Qual = $-1$. They have highly disturbed stellar kinematics that make their models unreliable. We also gave $\rm Qual = 0$ to the galaxies whose models did not match well with the observed two-dimensional stellar kinematics. These galaxies are mostly low-mass (or low-$\sigma_{\rm e}$) ones (see \citetalias{Zhu2023a}, fig.~8). This may introduce some biases in our results due to the sample selection. However, we can still estimate some zeroth-order quantities, such as the mass and $M/L$, for the $\rm Qual = 0$ galaxies. These quantities do not depend on the quality of the data as much as the density slope or dark matter fraction. To check for possible biases, we have also computed the results for the zeroth-order quantities of the $\rm Qual \geqslant 0$ sample (9360 galaxies) and presented them in \aref{appendix:Qual0}.

\subsection{Jeans Anisotropic Modelling (JAM)}
\label{sec:jam}
In \citetalias{Zhu2023a}, we perform Jeans Anisotropic Modelling \citep[JAM;][]{Cappellari2008,Cappellari2020} to construct dynamical models for the whole sample. In this section, we only give a brief introduction to the modelling approach, and we refer readers to \citetalias{Zhu2023a} for more details. The JAM model allows for anisotropy in second velocity moments and two different assumptions on the orientation of velocity ellipsoid, i.e. $\rm JAM_{cyl}$ (cylindrically-aligned) and $\rm JAM_{sph}$ (spherically-aligned). The total mass model has three components: the nuclear supermassive black hole, stellar mass distribution, and dark matter mass distribution. The black hole mass is estimated from $M_{\rm BH}-\sigma_{\rm c}$ relation \citep{McConnell2011}, where $\sigma_{\rm c}$ is computed as mean stellar velocity dispersion within 1 FWHM of MaNGA PSF. For the stellar component, we use the Multi-Gaussian Expansion \citep[MGE;][]{Emsellem1994,Cappellari2002} method to fit SDSS r-band images and obtain the surface brightness. Then the surface brightness is deprojected to obtain the luminosity density of the kinematic tracer. The total density derived by the model is a robust quantity, independent of possible gradients in the stellar mass-to-light ratio ($M/L$). However, the decomposition of the total density into luminous and dark matter relies on an adopted stellar $M/L$. In this paper we assume the stellar $M/L$ to be constant, within the region where we have kinematics, to measure dark matter. The dark matter component is characterized by various assumptions: the mass-follows-light model which assumes that the total mass density traces the luminosity density (hereafter MFL model), the model which assumes a spherical NFW \citep{Navarro1996} dark halo (hereafter NFW model), the fixed NFW model which assumes a spherical NFW halo predicted by the stellar mass-to-halo mass relation in \citet{Moster2013} and mass-concentration relation in \citet{Dutton2014} (hereafter fixed NFW model), the model which assumes a generalized NFW \citep{Wyithe2001} dark halo (hereafter gNFW model). The gNFW profile is written as
\begin{equation}
    \rho_{_{\rm DM}}(r) = \rho_s\left(\frac{r}{r_s}\right)^{\gamma}\left(\frac{1}{2}+\frac{1}{2}\frac{r}{r_s}\right)^{-\gamma-3},
\end{equation}
where $r_s$ is the characteristic radius, $\rho_s$ is the characteristic density, and $\gamma$ is the inner density slope. For $\gamma=-1$, this function reduces to the NFW profile.

In this work, all JAM-inferred quantities of different models are taken from \citetalias{Zhu2023a}. We calculate the size parameters $R_{\rm e}$, $R_{\rm e}^{\rm maj}$, and $r_{1/2}$ from MGE models in SDSS r-band, and then scale the $R_{\rm e}$ and $R_{\rm e}^{\rm maj}$ by a factor of 1.35 following \citet{Cappellari2013a}. Here, $R_{\rm e}$ is the circularized half-light radius (effective radius), $R_{\rm e}^{\rm maj}$ is the semi-major axis of half-light elliptical isophote, and $r_{1/2}$ is the 3D half-light radius. We also derive the total r-band luminosity $L$ from the MGE models and correct for the dust extinction effects (see \autoref{sec:sps} for details of the dust correction) to obtain the intrinsic luminosity. All quantities related to the luminosity, e.g. the total mass-to-light ratios, have been corrected for dust extinction. The velocity dispersion within an elliptical half-light isophote (with an area of $\pi R_{\rm e}^2$) is defined as 
\begin{equation}\label{eq:sigmae}
    \sigma_{\rm e} \approx \langle v_{\rm rms}^2\rangle_{\rm e}^{1/2} = \sqrt{\frac{\sum_k F_k (V_k^2+\sigma_k^2)}{\sum_k F_k}},
\end{equation}
where $F_k$, $V_k$, and $\sigma_k$ are the flux, stellar velocity, and stellar velocity dispersion in the $k$-th IFU spaxel. We define $M_{1/2}$ as the enclosed total mass within a sphere of $r_{1/2}$, which is derived from the best-fitting JAM model. The $(M/L)_{\rm e}$ is the total (dark plus luminous) mass-to-light ratio within a sphere of 1$R_{\rm e}$, which is derived from the JAM models with a dark matter halo (e.g. NFW models or gNFW models). We showed in \citetalias{Zhu2023a} that the total mass-to-light ratio $(M/L)_{\rm JAM}$ measured with MFL models is generally highly consistent with the integrated value $(M/L)_{\rm e}$ but gives a more robust estimate of the true total $M/L$ when the data have lower quality. To avoid confusion on the two expressions, we only use $(M/L)_{\rm JAM}$ to represent the total $M/L$ for all models in the following sections. The dynamical mass $M_{\rm JAM}$ is defined as
\begin{equation}
\label{eq:Mjam}
    M_{\rm JAM} \equiv (M/L)_{\rm JAM}\times L \approx M_{1/2}\times2.
\end{equation}
Since the systematic uncertainties in different models have been demonstrated to be small for $\rm Qual\geqslant1$ galaxies \citepalias{Zhu2023a}, we mainly use the NFW model with $\rm JAM_{cyl}$ as a reference model (if not mentioned otherwise) and another mass model as a comparison in the following sections. 

\subsection{Stellar Population Synthesis (SPS)}
\label{sec:sps}
The stellar population properties (e.g. age, metallicity, and stellar mass-to-light ratio) used in this paper are provided in \citetalias{Lu2023a}. We fit the IFU spectra of MaNGA DRP \citep{Law2016} data cubes using the \textsc{ppxf} software \citep{Cappellari2004,Cappellari2017,Cappellari2022} with the \textsc{fsps} models \citep{Conroy2009,Conroy2010}. Furthermore, we adopt the Padova stellar evolutionary isochrone \citep{Girardi2000} and the Salpeter \citep{Salpeter1955} initial mass function (IMF). We use 43 ages linearly spaced in $\rm \lg\,(Age/yr)$ between 6 and 10.2 (i.e. from 1 Myr to 15.85 Gyr) and 9 metallicities ($\mathrm{[Z/H]}=[-1.75,\,-1.5,\,-1.25,\,-1,\,-0.75,\,-0.5,\,-0.25,\,0,\,0.25]$). We correct for the dust extinction effects of the Milky Way (MW) and the observed galaxy itself, using the two-steps procedure briefly described here: we correct for the MW extinction by assuming \citet{Calzetti2000} extinction curve and adopting the $\mathrm{E(B-V)}$ values from the Galactic dust extinction map \citep{Schlegel1998}, then we perform \textsc{ppxf} fitting on the MW-extinction corrected spectrum and evaluate the dust extinction of target galaxies by setting the \texttt{dust} keyword in the updated \textsc{ppxf} software \citep{Cappellari2022} with assuming a two-parameter attenuation function f($A_{\rm v},\delta$). Here, $A_{\rm v}$ is the attenuation and $\delta$ is the UV slope at V-band ($\lambda=5500 \Angstrom$). More details of the dust correction can be found in \citetalias{Lu2023a} and \citet{Cappellari2022}.

We calculate the luminosity weighted $\lg \rm Age$ and metallicity $\mathrm{[Z/H]}$ using
\begin{equation}
    x = \frac{\sum^{N}_{i=1}w_{i}L_{i}x_{i}}{\sum^{N}_{i=1}w_{i}L_{i}},
\end{equation}
where $w_{i}$ is the weight of the $i$-th template, $L_{i}$ is the SDSS r-band luminosity of the $i$-th template, and $x_{i}$ is the $\rm \lg Age$ (or $\mathrm{[Z/H]}$) of the $i$-th template. Similarly, the stellar mass-to-light ratio is calculated as 
\begin{equation}
    (M_{\ast}/L)_{\rm SPS}=\frac{\sum_{i=1}^{N} w_{i} M_{i}^{\rm nogas}}{\sum_{i=1}^{N} w_{i} L_{i}},
\end{equation}
where $M_{i}^{\rm nogas}$ is the stellar mass of the $i$-th template, which includes the mass of living stars and stellar remnants but excludes the mass of lost gas during stellar evolution. To obtain the global properties and their gradients, the spectra are stacked in two ways: (1) The spectra within the elliptical half-light isophote are stacked to obtain a spectrum with high signal-to-noise ratio (fig.~4 in \citetalias{Lu2023a}), then we fit the stacked spectrum to obtain the global stellar population properties. (2) The spectra are Voronoi binned \citep{Cappellari2003} to $\rm S/N=30$, then we fit the stacked spectrum in each bin and finally obtain a map of stellar population for each galaxy. We estimate the stellar mass 
\begin{equation}
\label{eq:MsSPS}
M_{\ast} = (M_{\ast}/L)_{\rm SPS}(<R_{\rm e}) \times L,
\end{equation}
where $(M_{\ast}/L)_{\rm SPS}(<R_{\rm e})$ is the r-band stellar mass-to-light ratio derived from the stacked spectrum within elliptical half-light isophote and $L$ is the r-band total luminosity derived from MGE models. The stellar population gradients are calculated by linearly fitting the stellar population profile within an effective radius (see the details about the calculation in \citetalias{Lu2023a}).

\begin{figure}
    \centering
    \includegraphics[width=\columnwidth]{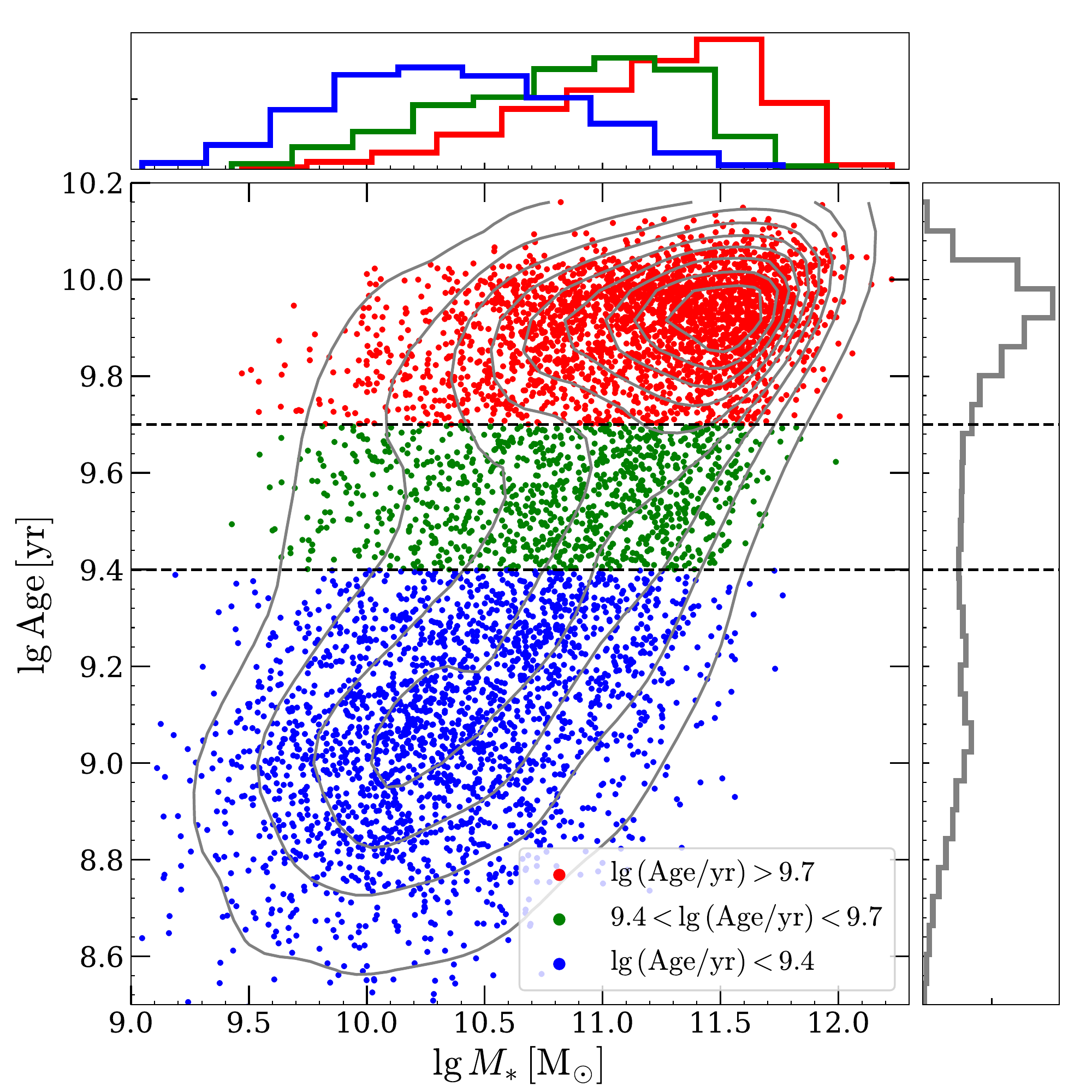}
    \caption{The Age - $M_{\ast}$ plane for the $\rm Qual\geqslant1$ galaxies, with coloured symbols representing the galaxies in different stellar age bins (red: $\lg({\rm Age/yr})>9.7$, green: $9.4<\lg({\rm Age/yr})<9.7$, blue: $\lg({\rm Age/yr})<9.4$). The stellar mass values are taken from the SPS models (see \autoref{eq:MsSPS}), which assume a Salpeter IMF \citep{Salpeter1955}. The grey contours show the kernel density estimate for the two-dimensional galaxy distribution (using \href{https://docs.scipy.org/doc/scipy/reference/generated/scipy.stats.gaussian_kde.html}{scipy\_stats\_gaussian\_kde}). Histograms show the probability density functions (normalised to unity) for the galaxies in different stellar age bins (red, green, and blue) and full sample (grey).} 
    \label{fig:Age_Ms}
\end{figure}

Based on the stellar age, we split the galaxies into old, intermediate, and young galaxies, using the selection as follows:
\begin{itemize}
    \item Old: $\rm \lg(Age/yr)>9.7$
    \item Intermediate: $\rm 9.4<\lg(Age/yr)<9.7$
    \item Young: $\rm \lg(Age/yr)<9.4$
\end{itemize}
Under these selection criteria, there are 2734 old galaxies, 1019 intermediate galaxies, and 2199 young galaxies. In \autoref{fig:Age_Ms}, we present the bi-modal galaxy distribution in the $\rm Age-M_{\ast}$ diagram, which suggests that the classification based on stellar age qualitatively (but not strictly) corresponds to the classification (i.e. the red sequence, blue cloud, and green valley) based on colour-magnitude diagram \citep{Strateva2001,Bell2003}.

\subsection{The morphology, environment and stellar angular momentum}
\label{sec:morph_env_lambda}
We divide the whole MaNGA sample into ETGs and LTGs. The ETGs include elliptical (E) and lenticular (S0) galaxies, while the LTGs correspond to spiral (S) galaxies. The classification of morphology is based on the MaNGA Deep Learning (DL) morphological catalogue \citep{Dominguez-Sanchez2022}. To obtain the most clean morphological samples, we use the most restrictive selection \citep[being recommended in sec.~3.4.1 of][]{Dominguez-Sanchez2022} as follows:
\begin{itemize}
    \item E: ($\rm P_{LTG}<0.5$) and (T-Type<0) and ($\rm P_{S0}<0.5$) and ($\rm VC=1$) and ($\rm VF=0$)
    \item S0: ($\rm P_{LTG}<0.5$) and (T-Type<0) and ($\rm P_{S0}>0.5$) and ($\rm VC=2$) and ($\rm VF=0$)
    \item S: ($\rm P_{LTG}>0.5$) and (T-Type>0) and ($\rm VC=3$) and ($\rm VF=0$)
\end{itemize}
This selection combines the information of three classification models provided in \citet{Dominguez-Sanchez2022}: 1) the T-Type values; 2) the two binary classifications: the $\rm P_{LTG}$ separates ETGs from LTGs and the $\rm P_{S0}$ separate S0s from Es; 3) the visual classification: VC (1 for Es, 2 for S0s, 3 for Ss) and VF (0 for certain visual classification, 1 for uncertain visual classification). For the $\rm Qual \geqslant 1$ galaxies, this selection returns 1621 Es, 603 S0s, 2966 Ss, and 925 unclassified galaxies, respectively.

We match the MaNGA galaxies to the group catalogue derived by \citet{Yang2007} (hereafter Yang07), which uses an adaptive halo-based group finder to assign each galaxy in the SDSS DR7 \citep{SDSSDR7} sample to a group. For each group, the galaxy with the largest stellar mass is assumed to be the central galaxy, while others are assumed to be satellite galaxies. By finding the MaNGA galaxies' counterparts in the Yang07 catalogue, we classify the $\rm Qual \geqslant 1$ galaxies into 4081 central and 1052 satellite galaxies. 

\begin{figure}
    \centering
    \includegraphics[width=\columnwidth]{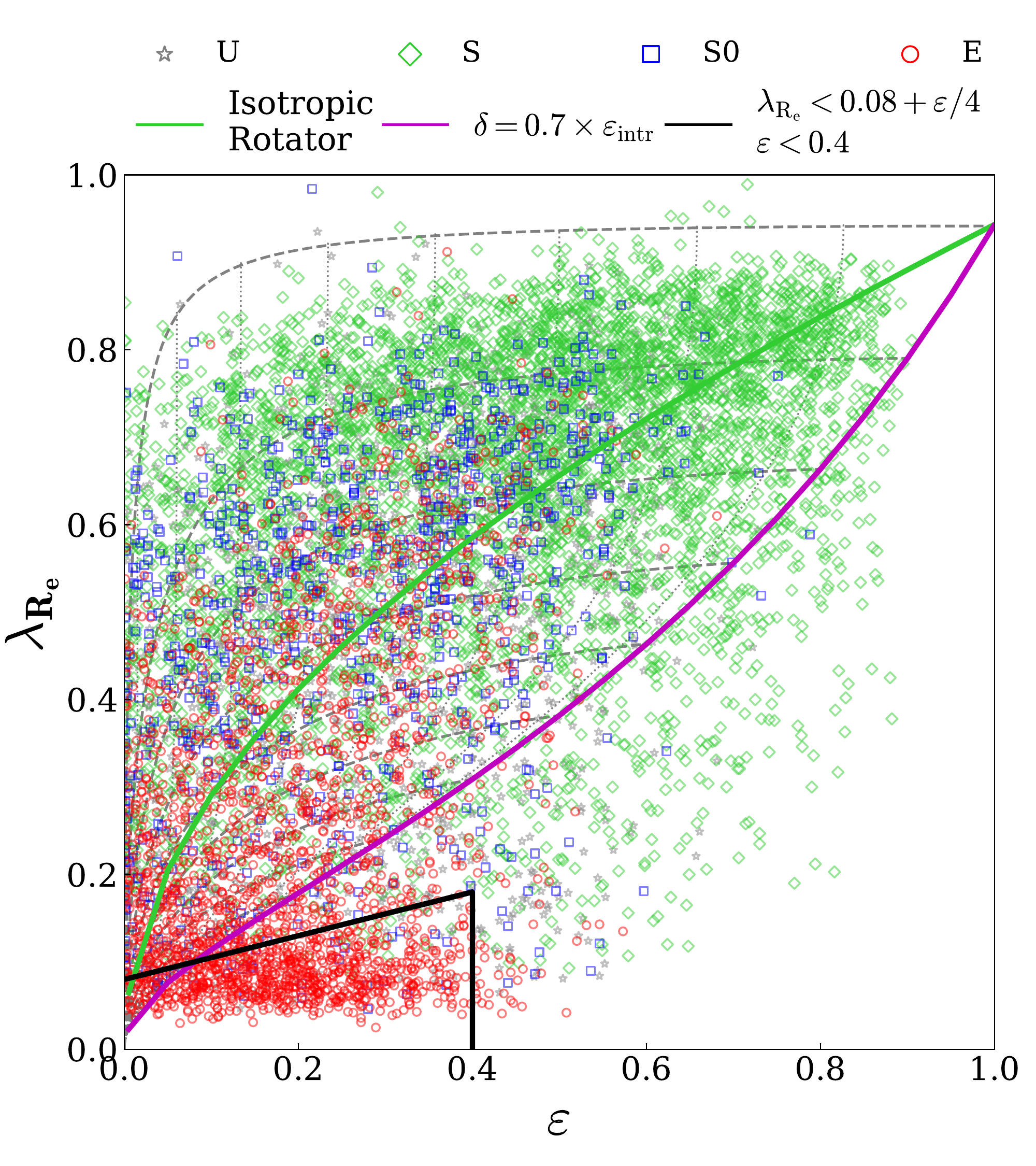}
    \caption{The $(\lambda_{\rm R_{\rm e}},\varepsilon)$ diagram, where $\lambda_{\rm R_{\rm e}}$ is the beam corrected spin parameter and $\varepsilon$ is the observed ellipticity derived from MGE fitting. The galaxies are classified as elliptical (E), spiral (S), and lenticular (S0) galaxies, corresponding to the red, cyan, and blue symbols in the diagram. The galaxies with unclassified morphology are shown as grey symbols. The green line represents the predicted relation for an edge-on ($i=90^{\circ}$) isotropic rotator \citep[][eq. 14]{Binney2005,Cappellari2016ARAA}, while the magenta line denotes the edge-on shape-anisotropy upper-limit from \citet{Cappellari2007} and \citet[][eq. 11]{Cappellari2016ARAA}. The thin dotted lines show how the magenta line changes with different inclinations ($\Delta i=10^{\circ}$), while the thick dashed lines show how the galaxies move across the diagram with changing inclination for a set of given $\varepsilon_{\rm intr}$ values ($\Delta \varepsilon_{\rm intr}=0.1$). The lower-left region enclosed by the black solid lines ($\lambda_{\rm R_{\rm e}}<0.08+\varepsilon/4,\varepsilon<0.4$) defines the region occupied by slow rotators \citep[][eq. 19]{Cappellari2016ARAA}.}
    \label{fig:eps_lambda}
\end{figure}

The proxy for stellar angular momentum (or spin parameter) $\lambda_{\rm R_e}$ is defined within the same aperture as $\sigma_{\rm e}$ (i.e. elliptical half-light isophote), written as \citep{Emsellem2007}
\begin{equation}\label{eq:lambdaRe}
    \lambda_{R_{\rm e}} = \frac{\sum_k F_k R_k|V_k|}{\sum_k F_k R_k \sqrt{V_k^2+\sigma_k^2}},
\end{equation}
where $F_k$, $V_k$ and $\sigma_k$ are the same as \autoref{eq:sigmae}; $R_k$ is the distance of $k$-th spaxel to the galaxy centre. The $\lambda_{\rm R_e}$ has been corrected for the beam smearing effect following \citet{Graham2018}\footnote{\url{https://github.com/marktgraham/lambdaR_e_calc}}. In \autoref{fig:eps_lambda}, we show the ($\lambda_{\rm R_e}$, $\varepsilon$) diagram for 9360 $\rm Qual \geqslant 0$ galaxies classified as Es, S0s, and Ss, where $\varepsilon$ is the observed ellipticity within the half-light isophote derived from MGE models using the \textsc{mge\_half\_light\_isophote} procedure in the \textsc{JamPy} package\footnote{Version 6.3.3, available from \url{https://pypi.org/project/jampy/}}. We define the slow rotators (SRs) as the galaxies satisfying $\lambda_{\rm R_e}<0.08+\varepsilon/4$ and $\varepsilon<0.4$ (the region enclosed by black solid lines in \autoref{fig:eps_lambda}) following \citet[equation~19]{Cappellari2016ARAA}, while the fast rotators (FRs) are defined to be the galaxies outside the region occupied by the SRs. Under this definition, the $\rm Qual \geqslant 1$ galaxies used in this paper consist of 639 SRs and 5426 FRs. Among the 639 SRs, there are 592 Es, 23 S0s, and 6 Ss.

\section{Results}
\label{sec:results}
\subsection{The fundamental plane (FP) and mass plane (MP)}
\label{sec:FP_MP}
In this section, we present the FP and MP, which are obtained using the \textsc{lts\_planefit}\footnote{Available from \url{https://pypi.org/project/ltsfit/}} software \citep{Cappellari2013a}. The \textsc{lts\_planefit} procedure combines the Least Trimmed Squares robust technique of \citet{Rousseeuw2006} with a least-squares fitting algorithm, allowing for the errors in all variables and the intrinsic scatter. In the fitting, we adopt 10 per cent error of luminosity $L$, 5 per cent error of $\sigma_{\rm e}$, and 10 per cent error of $R_{\rm e}$ \citep{Cappellari2013a}, while the error of $M_{1/2}$ is 12 per cent \citepalias{Zhu2023a}. During the \textsc{lts\_planefit} fitting, we set the sigma-clipping keyword \texttt{clip=4} to avoid removing too many galaxies.

\begin{figure*}
    \centering
    \includegraphics[width=0.85\textwidth]{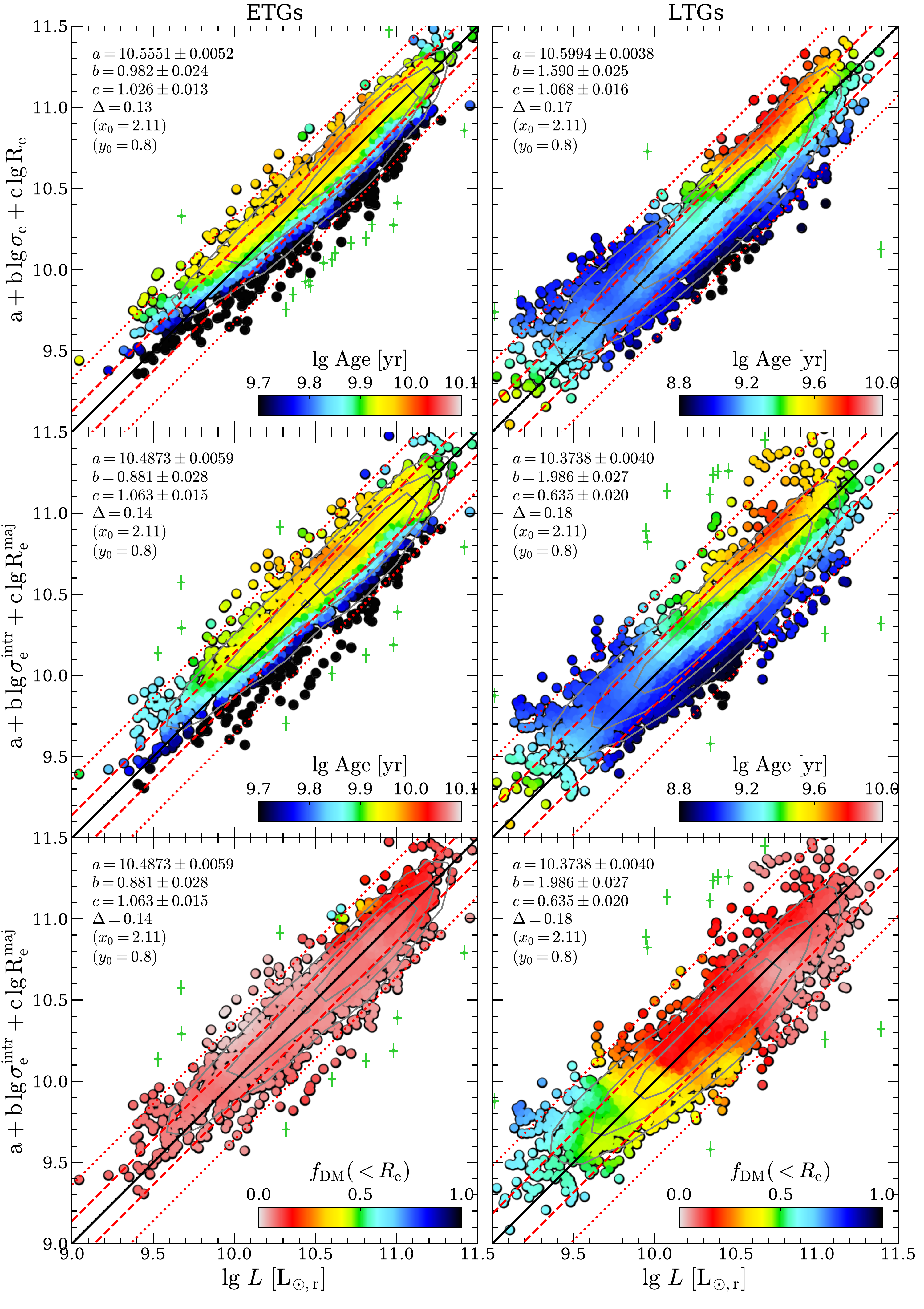}
    \caption{The fundamental plane for early-type (left column) and late-type (right column) galaxies, with colours in each panel showing the \textsc{loess}-smoothed (\texttt{frac=0.1}) $\rm \lg Age$ and $f_{\rm DM}(<R_{\rm e})$ of corresponding galaxy sample. \emph{Top Panels}: Edge-on view of the FP. At each panel, the coefficients of the best-fitting plane $\rm z=a+b(x-x_0)+c(y-y_0)$ and the observed rms scatter $\Delta$ are obtained from the \textsc{lts\_planefit} procedure (with \texttt{clip=4}) and shown on the upper left corner. The black solid, red dashed, and red dotted lines represent the best-fitting, 1$\sigma$ error (68 per cent), and 2.6$\sigma$ error (99 per cent), respectively. The symbols within 4$\sigma$ error are coloured by stellar age, while the green crosses are the outliers beyond 4$\sigma$ error. The grey contours show the $1\sigma$, $2\sigma$ and $3\sigma$ confidence level of the two-dimensional distribution. \emph{Middle Panels}: The Symbols, colours, and lines are the same as in the top panels, but using the major axis $R_{\rm e}^{\rm maj}$ of the effective isophote and the deprojected second velocity moment $\sigma_{\rm e}^{\rm intr}$. \emph{Bottom panels}: The same as middle panels, but coloured by $f_{\rm DM}(<R_{\rm e})$.}
    \label{fig:FP}
\end{figure*}

In \autoref{fig:FP}, we present the FP, which is written as
\begin{equation}
\label{eq:FP}
    \lg\left(\frac{L}{{\rm L_{\odot,r}}}\right) = a+b\lg\left(\frac{\sigma_{\rm e}}{\rm km\,s^{-1}}\right)+c\lg\left(\frac{R_{\rm e}}{\rm kpc}\right),
\end{equation}
for the ETGs and LTGs. As opposed to the classic form of the FP using $\Sigma_{\rm e}=L/\pi R_{\rm e}^2$, the use of $L$ instead of $\Sigma_{\rm e}$ reduces the covariance between $\Sigma_{\rm e}$ and $R_{\rm e}$, due to the fact that using $\Sigma_{\rm e}$ the radius would appear on both axes. For the ETGs shown in the top left panel of \autoref{fig:FP}, the coefficients and rms scatter of the FP are $b=0.982$, $c=1.026$, and $\Delta=0.13$ dex (35 per cent), which are similar to those of the FPs derived from $\rm ATLAS^{3D}$ \citep[$b=1.249$, $c=0.964$, and $\Delta=0.10$ dex;][]{Cappellari2013a} and $\rm SAMI$ \citep[$b=1.294$, $c=0.912$, and $\Delta=0.104$ dex;][]{DEugenio2021}. The FPs for cluster member ETGs show similar coefficients, e.g. $b=0.89$, $c=0.95$, and $\Delta=0.07$ dex in \citet{Scott2015}; $b=1.03$, $c=1.07$, and $\Delta=0.087$ dex in \citet{Shetty2020}. In agreement with previous studies on the FP of ETGs as mentioned above, our $b$ and $c$ values are inconsistent with the expected coefficients from the scalar virial equation (i.e. $b=2$ and $c=1$). We also investigate the FP of LTGs in the top right panel of \autoref{fig:FP}, resulting in the different coefficients ($b=1.590$, $c=1.068$) and a larger rms scatter ($\Delta=0.17$ dex). The differences in the FP between ETGs and LTGs may be due to the LTGs' rotation-supported kinematics and disk-like stellar component, which has greater projection effects on the measurements of $\sigma_{\rm e}$ and the $R_{\rm e}$.

Following \citet{Cappellari2013a}, we use the $R_{\rm e}^{\rm maj}$ and the deprojected second velocity moment $\sigma_{\rm e}^{\rm intr}$ instead of the $R_{\rm e}$ and the $\sigma_{\rm e}$ to reduce the effect of inclination. Given that the velocity ellipsoid in ETGs is generally close to a sphere \citep{Gerhard2001,Cappellari2007,Thomas2009} and the kinematics are dominated by rotation in LTGs, we suppose that the velocity dispersion changes weakly with inclination, while the light-of-sight velocity varies as $V=v/\sin{i}$, where $i$ is the inclination inferred from NFW models and $v$ is the velocity being edge-on ($i=90^{\circ}$). Thus the deprojected second velocity moment is defined as 
\begin{equation}
    \sigma_{\rm e}^{\rm intr} \approx \langle v_{\rm rms}^2\rangle_{\rm e, intr}^{1/2} = \sqrt{\frac{\sum_k F_k (V_k^2/\sin^2i+\sigma_k^2)}{\sum_k F_k}},
\end{equation}
where $F_k$, $V_k$, and $\sigma_k$ are the flux, light-of-sight stellar velocity, stellar velocity dispersion in the $k$-th IFU spaxel, $i$ is the inclination derived from best-fitting JAM models. We found that the observed rms scatters between $\sigma_{\rm e}$ and $\sigma_{\rm e}^{\rm intr}$ are $\Delta=0.053$ dex (13 per cent) for ETGs and $\Delta=0.074$ dex (19 per cent) for LTGs, which are larger than the random error $\Delta=0.025$ dex derived in \citet{Cappellari2013a}. The deprojected FP ($L$, $\sigma_{\rm e}^{\rm intr}$, $R_{\rm e}^{\rm maj}$) for ETGs has the nearly unchanged coefficients ($b=0.881$ and $c=1.063$), while the coefficients for LTGs significantly change to $b=1.986$ and $c=0.635$ (the middle panels of \autoref{fig:FP}). Furthermore, the rms scatters of the deprojected FPs ($\Delta=0.14$ dex for ETGs and $\Delta=0.18$ dex for LTGs) remain nearly the same as the FPs, suggesting that the scatter of the FP is not driven by the projection effects.

To explore the origin of the FP scatter, the FPs in \autoref{fig:FP} are coloured by the luminosity-weighted stellar age, which is smoothed using the locally-weighted regression method by \citet{Cleveland1988} as implemented by \citet{Cappellari2013b} in the {\sc loess}\footnote{Available from \url{https://pypi.org/project/loess/}} software (unless otherwise specified, we adopt a small \texttt{frac=0.05} throughout this paper to avoid over-smoothing, given the large number of values in our sample). A two-dimensional {\sc loess}-smoothed map is a way of showing the average value of a function that depends on two variables. It is the two-dimensional analogue of the average trend that is often shown in one-dimensional plots. For the ETGs, the variation in age shows a strong trend perpendicular to the FP, in agreement with the trends found in the nearby galaxies from the SDSS survey \citep[fig.~7]{Graves2009}, the SAMI survey \citep[fig.~9]{DEugenio2021} and the galaxies of LEGA-C survey at $0.6<z<1$ \citep[fig.~6]{deGraaff2021}. A similar trend is also found in the $L>10^{10.2}{\rm L_{\odot,r}}$ LTGs, but no correlation between the age and the residuals of the FP is observed for the less luminous LTGs. Since the stellar age correlates to the stellar mass-to-light ratio $M_{\ast}/L$, the $M_{\ast}/L$ is probably the driving mechanism for the scatter of the FP. We confirm this in the comparisons between the FP and the MP.

\begin{figure*}
    \centering
    \includegraphics[width=0.85\textwidth]{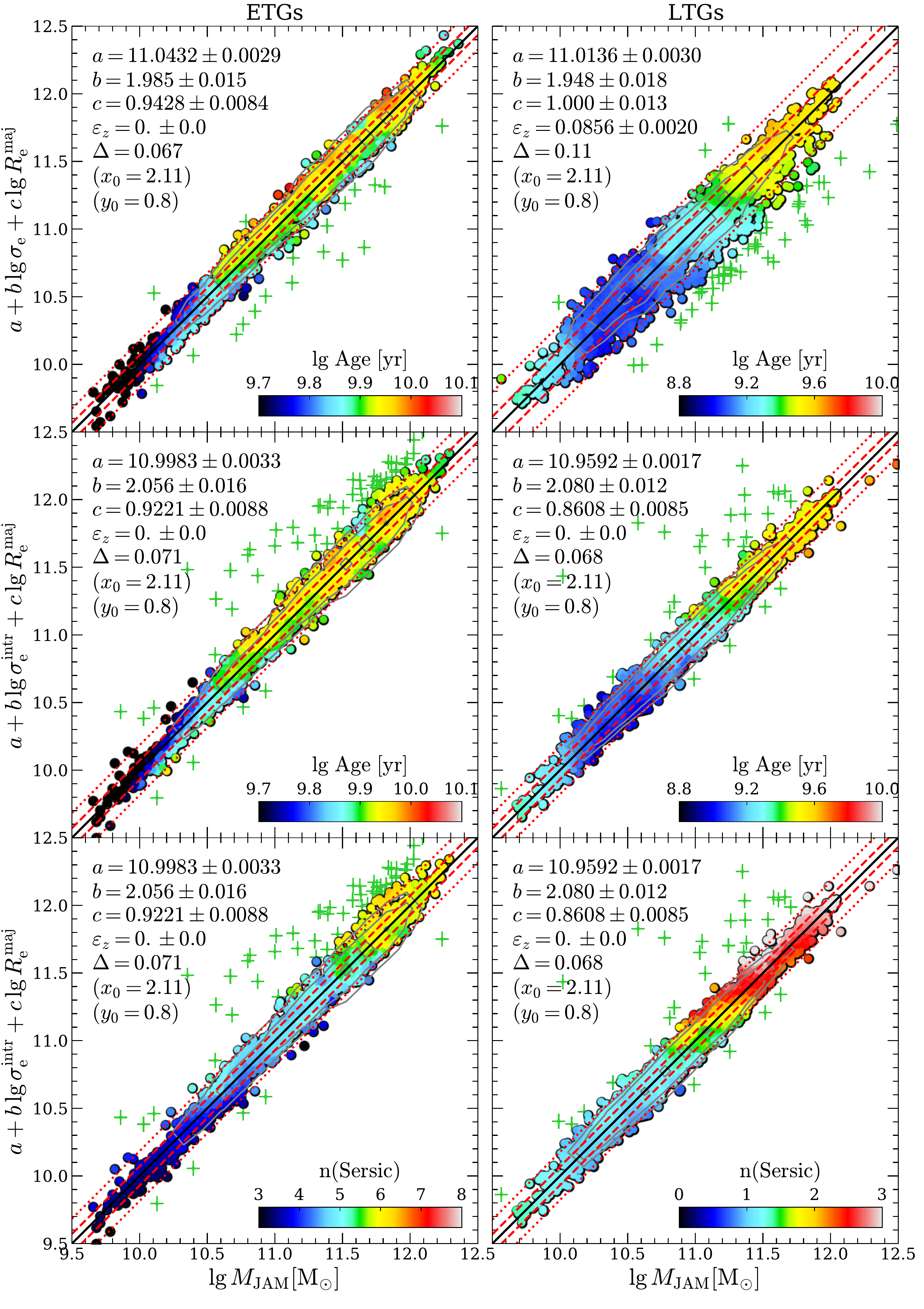}
    \caption{The mass planes for early-type (left column) and late-type (right column) galaxies, with colours in each panel showing the {\sc loess}-smoothed (\texttt{frac=0.1}) $\rm \lg Age$ and Sersic index $n$ of corresponding galaxy sample. The panels are similar to those in \autoref{fig:FP}, but substituting the total luminosity $L$ with the JAM inferred total mass $M_{\rm JAM} \equiv (M/L)_{\rm JAM}\times L$ (MFL models). Furthermore, the MPs in the top panels use $R_{\rm e}^{\rm maj}$ instead of $R_{\rm e}$ following previous studies \citep{Cappellari2013a,Lihongyu2018,Shetty2020}.}
    \label{fig:MP}
\end{figure*}

We replace the r-band total luminosity $L$ with the dynamical mass derived from MFL models, $M_{\rm JAM} \equiv (M/L)_{\rm JAM}\times L$, to obtain the MP in the form of
\begin{equation}
\label{eq:MP}
    \lg\left(\frac{M_{\rm JAM}}{{\rm M_{\odot}}}\right) = a+b\lg\left(\frac{\sigma_{\rm e}}{\rm km\,s^{-1}}\right)+c\lg\left(\frac{R_{\rm e}^{\rm maj}}{\rm kpc}\right).
\end{equation}
Note that the $R_{\rm e}$ is also replaced with $R_{\rm e}^{\rm maj}$ to reduce the projection effects on the size, following \citet{Cappellari2013a,Lihongyu2018,Shetty2020}. The MPs for ETGs and LTGs are shown in the top panels of \autoref{fig:MP}. In agreement with previous studies \citep{Cappellari2013a,Lihongyu2018,Shetty2020,DEugenio2021}, we find that the coefficients of the MP ($b=1.985$ and $c=0.9428$ for ETGs, $b=1.948$ and $c=1.000$ for LTGs) become much closer to the virial one ($b=2$ and $c=1$). The observed scatters also significantly decrease (from $\Delta=0.13$ dex to $\Delta=0.067$ dex for ETGs, from $\Delta=0.17$ dex to $\Delta=0.11$ dex for LTGs), resulting in the negligible intrinsic scatter for ETGs ($\varepsilon_{\rm z}=0$) and the significant intrinsic scatter for LTGs ($\varepsilon_{\rm z}=0.0856$ dex). This confirms previous findings that much of the tilt and the scatter of the FP is due to the variations in dynamical $M/L$ along and perpendicular to the FP for ETGs \citep{Cappellari2006,Cappellari2013a,Bolton2008,Auger2010,Thomas2011,deGraaff2021,DEugenio2021}.

However, the much larger intrinsic scatter of the MP for LTGs indicates another driving mechanism, which is likely to be the projection effects as discussed above. Thus we show the deprojected MPs ($M_{\rm JAM}$, $\sigma_{\rm e}^{\rm intr}$, $R_{\rm e}^{\rm maj}$) in the middle panels of \autoref{fig:MP}, which are derived by replacing $\sigma_{\rm e}$ with the deprojected velocity second moment $\sigma_{\rm e}^{\rm intr}$. As expected, the deprojected MP of ETGs remains nearly unchanged (both the coefficients and the scatter), while the coefficients of the deprojected MP for LTGs become slightly closer to the virial predictions. A remarkable finding is the significant decrease in the scatters (both observed and intrinsic) of the deprojected MP for LTGs, resulting in the scatters that are comparable with ETGs' ($\Delta=0.071$ dex and $\varepsilon_{\rm z}=0$ for ETGs, $\Delta=0.068$ dex and $\varepsilon_{\rm z}=0$ for LTGs). The intrinsic scatters $\varepsilon_{\rm z}$ remain zero until we reduce the errors of $M_{\rm JAM}$ to be 5 per cent for ETGs and 10 per cent for LTGs, while keeping 5 per cent errors of $\sigma_{\rm e}$ and 10 per cent errors of $R_{\rm e}$. The very small intrinsic scatter ($\varepsilon_{\rm z}=0$), as well as the invisible variation of \citet{Sersic1968} index perpendicular to the MP (bottom panels in \autoref{fig:MP}), confirm the negligible contribution of structural non-homology (captured by the Sersic index) to the scatter of the FP \citep{Cappellari2006,Cappellari2013a,Bolton2008,Auger2010,deGraaff2021}. However, \citet{DEugenio2021} also found that non-homology accounts for $\sim20$ per cent of the FP scatter for the SAMI ETGs. The discrepancy is likely due to the non-negligible scatter of the virial mass estimator \citep[][fig.~7]{Cappellari2006,vanderWel2022} that they adopted to estimate the dynamical masses. It is worth mentioning that we assume a spherical dark matter halo and an axisymmetric oblate stellar component in our dynamical models, which limits the range of homology violations that the models can represent. To investigate more effects of structural non-homology (e.g. the dark matter halo shape, the triaxiality of the stellar system) would require more general dynamical models, which is beyond the scope of this paper.

As opposed to the FP, the variation of age perpendicular to the MP is not observed. For the ETGs, the result clearly shows that the scatter of the FP is mainly due to the variation in stellar mass-to-light ratio, as the dark matter fraction is generally small (see \autoref{fig:fdm_Ms_type}). The trend is also found in the $L>10^{10.2}{\rm L_{\odot,r}}$ LTGs. However, the stellar mass-to-light ratio can not fully explain the scatter of the FP for LTGs, especially for the $L<10^{10.2}{\rm L_{\odot,r}}$ LTGs without age variation perpendicular to the FP (right panels in \autoref{fig:FP}). This implies that the scatter of the FP for these galaxies is dominated by the variation in dark matter fraction, confirmed by the bottom panels of \autoref{fig:FP}.

In summary, we come to three conclusions in this section: (i) The deprojected MPs for both ETGs and LTGs, which have been corrected for the projection effects, are very close to the virial predictions in the sense of both the coefficients ($b\approx2$ and $c\approx1$) and the scatter ($\Delta\approx0.06-0.07$ dex and $\varepsilon_{\rm z}=0$). The projection effects are stronger for the MP ($M_{\rm JAM}$, $\sigma_{\rm e}$, $R_{\rm e}^{\rm maj}$) of LTGs, while the projection effects are very weak for the MP of ETGs; (ii) The tilt and the scatter of the FP are mainly due to the variations of the total $M/L$ along and perpendicular to the FP, not to non-homology in light profiles; (iii) For ETGs, the variation in stellar mass-to-light ratio $M_{\ast}/L$ dominates the variation in total $M/L$ and further the scatter of FP. For LTGs, the scatter of FP is owing to the variation of $M_{\ast}/L$ for the luminous population ($L>10^{10.2}{\rm L_{\odot,r}}$), while the variation in dark matter fraction $f_{\rm DM}(<R_{\rm e})$ plays a more important role for the fainter population. In \aref{appendix:Qual0}, we plot the FP and MP while also including galaxies with $\rm Qual=0$, namely for all galaxies with $\rm Qual \geqslant 0$, and find that the results of this section still hold.


\subsection{The \texorpdfstring{$(M/L) - \sigma_{\rm e}$}{} relation}
\label{sec:ML}
\begin{figure}
    \centering
    \includegraphics[width=\columnwidth]{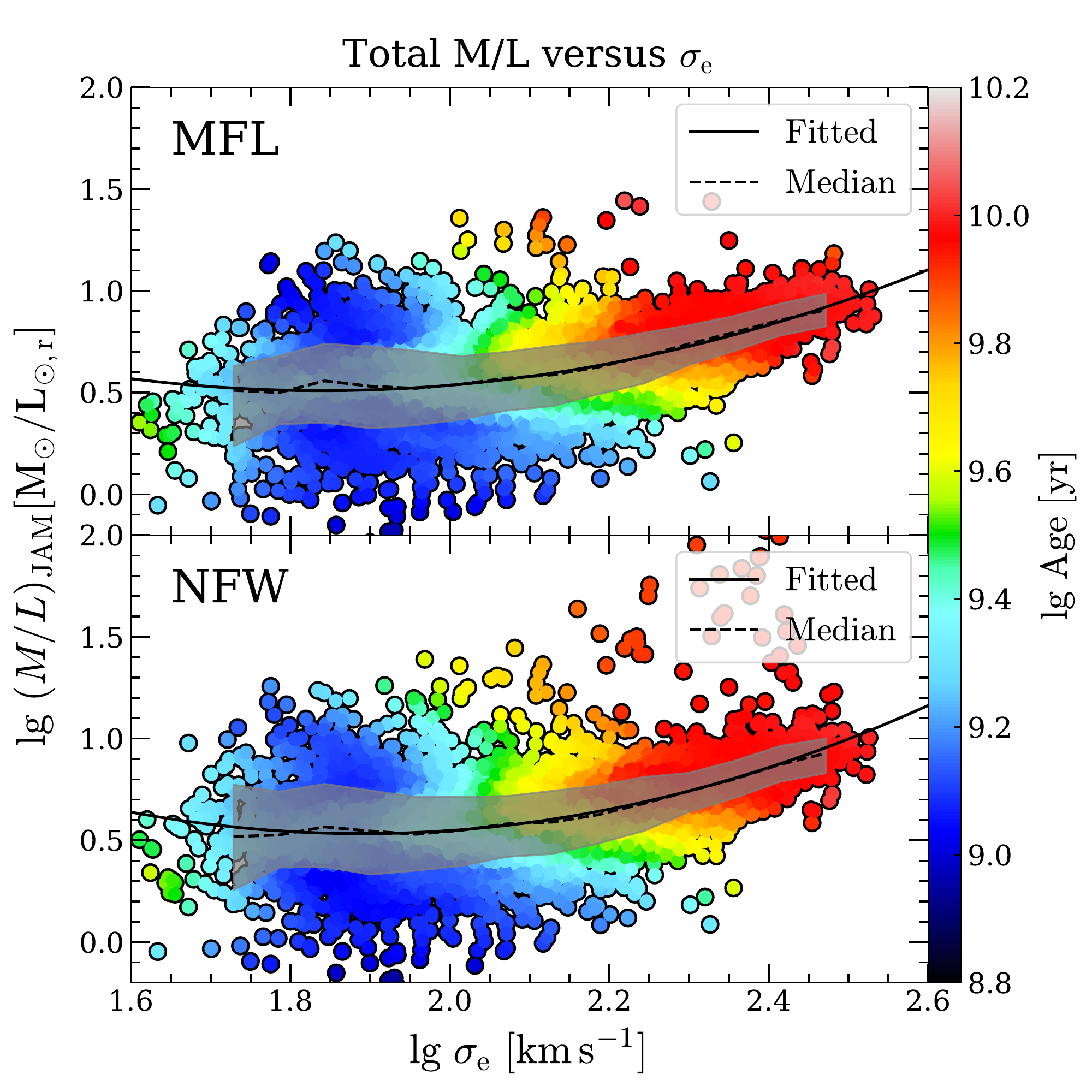}
    \caption{The dynamical mass-to-light ratios $(M/L)_{\rm JAM}$ as a function of $\sigma_{\rm e}$ for different mass models: mass-follows-light model (top panel) and NFW halo model (bottom panel). The symbols are coloured by stellar age using the \textsc{loess} software (\texttt{frac=0.05}). In each panel, the black dashed curve represents the median value, while the grey shaded region denotes the [16th, 84th ] percentile of values. The 1$\sigma$ errors range from 0.20 dex at low-$\sigma_{\rm e}$ end to 0.079 dex at high-$\sigma_{\rm e}$ end (a median value of 0.15 dex). The black solid curves are the best-fitting parabolic relations in the form of \autoref{eq:MLdyn_fit}, with $[\lg(M/L)_{0}, A, \lg\sigma_{0}] = [0.51, 1.03, 1.84]$ for the mass-follows-light model (top) and $[0.53, 1.25, 1.89]$ for the NFW model (bottom).}
    \label{fig:MLjam_sigma}
\end{figure}

\autoref{fig:MLjam_sigma} presents the $(M/L)_{\rm JAM}-\sigma_{\rm e}$ relations  \citep{Cappellari2006} for both MFL and NFW models. We find that the relations are quite similar between different models, and both of them can be well described using a parabolic relation
\begin{equation}
\label{eq:MLdyn_fit}
    \lg(M/L)_{\rm JAM} = \lg(M/L)_{0}+A\times(\lg \sigma_{\rm e}-\lg\sigma_{0})^2,
\end{equation}
with $[\lg(M/L)_{0}, A, \lg\sigma_{0}]=[0.51,1.03,1.84]$ for the MFL model and $[0.53,1.25,1.89]$ for the NFW model. The relations for a larger sample (i.e. $\rm Qual\geqslant0$) is presented in \aref{appendix:Qual0}, which are consistent with the parabolic relations (i.e. \autoref{eq:MLdyn_fit} for the $\rm Qual\geqslant1$ sample) at $\sigma_{\rm e} \gtrsim 60\,{\rm km\,s^{-1}}$. The unified relation is derived from various types of galaxies (including both ETGs and LTGs), extending the $(M/L)_{\rm JAM}-\sigma_{\rm e}$ relation to be more general than the linear relation adopted by previous studies who used dynamical models to measure $M/L$ as we did here \citep{Cappellari2006,Cappellari2013a,vanderMarel2007,Scott2015,Shetty2020}. Two features of this relation are obvious: (i) The $(M/L)_{\rm JAM}$ monotonically increases with increasing $\sigma_{\rm e}$, with a median 1$\sigma$ rms scatter (68 per cent) of $\approx 0.15$ dex; (ii) The slope and the scatter change with the $\sigma_{\rm e}$: the slope is steeper and the scatter is smaller for the galaxies with larger $\sigma_{\rm e}$ (0.20 dex at low-$\sigma_{\rm e}$ end and 0.079 dex at high-$\sigma_{\rm e}$ end).

\begin{figure}
    \centering
    \includegraphics[width=\columnwidth]{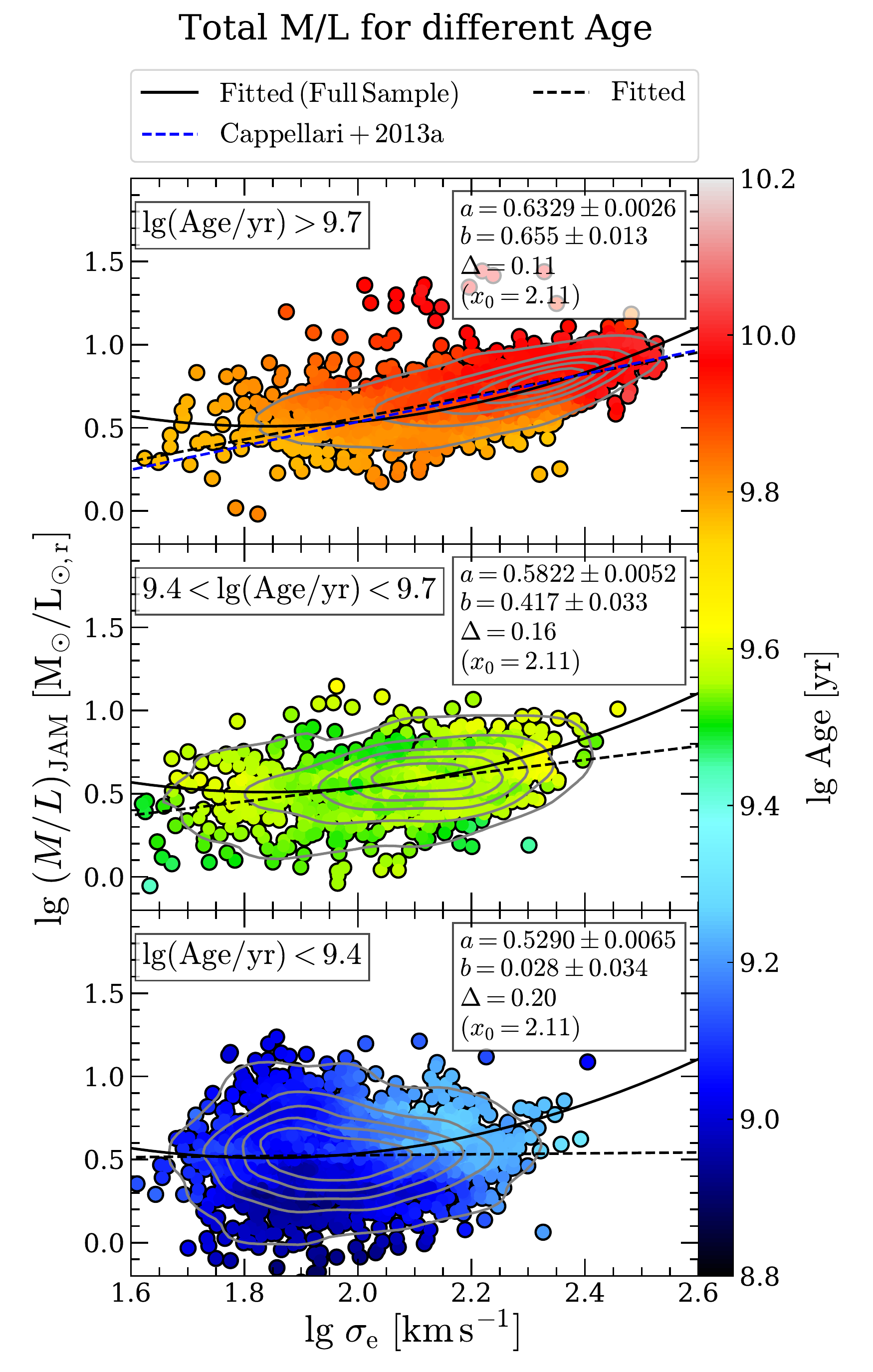}
    \caption{The mass-follows-light models inferred dynamical mass-to-light ratios as a function of $\sigma_{\rm e}$ for galaxies with different stellar ages. From top to bottom, the relations for old galaxies, intermediate galaxies, and young galaxies are presented. The grey contours are the kernel density estimate for the galaxy distribution. The symbols are coloured by stellar age. The black solid curve is the best-fitting relation derived from the full sample, while the black dashed lines(curves) represent the best-fitting relation of the corresponding subsample. The best-fitting straight line $y=a+b(x-x_{_{0}})$ is obtained using the {\sc lts\_linefit} procedure, with the coefficients shown in the corresponding panel. The relation obtained from $\rm ATLAS^{3D}$ \citep{Cappellari2013a} for ETGs is shown as the blue dashed line.}
    \label{fig:MLjam_sigma_agetype}
\end{figure}

In addition, we also find that the $(M/L)_{\rm JAM}-\sigma_{\rm e}$ relation is steeper and has a smaller scatter for the galaxies with older stellar age (the top panel of \autoref{fig:MLjam_sigma_agetype}). In \autoref{fig:MLjam_sigma_agetype}, we present the $(M/L)_{\rm JAM}-\sigma_{\rm e}$ relations for the galaxies within different stellar age bins. At each age bin, we obtain the best-fitting linear relation using the $\textsc{lts\_linefit}$ procedure \citep{Cappellari2013a}. The slopes of the $(M/L)_{\rm JAM}-\sigma_{\rm e}$ relations become steeper with increasing stellar age: the relation is nearly flat ($b=0.028$) for the youngest galaxy population and is steeper ($b=0.417$) for the older population and finally becomes the steepest ($b=0.655$) for the oldest population. Specifically, for the oldest galaxies (second panel in \autoref{fig:MLjam_sigma_agetype}), the coefficients ($a=0.6329$, $b=0.655$ and $\Delta=0.11$ dex) are quite similar to those found in the $\rm ATLAS^{3D}$ ETGs \citep[$a=0.6151$, $b=0.72$ and $\Delta=0.11$ dex;][]{Cappellari2013a}.

\begin{figure}
    \centering
    \includegraphics[width=\columnwidth]{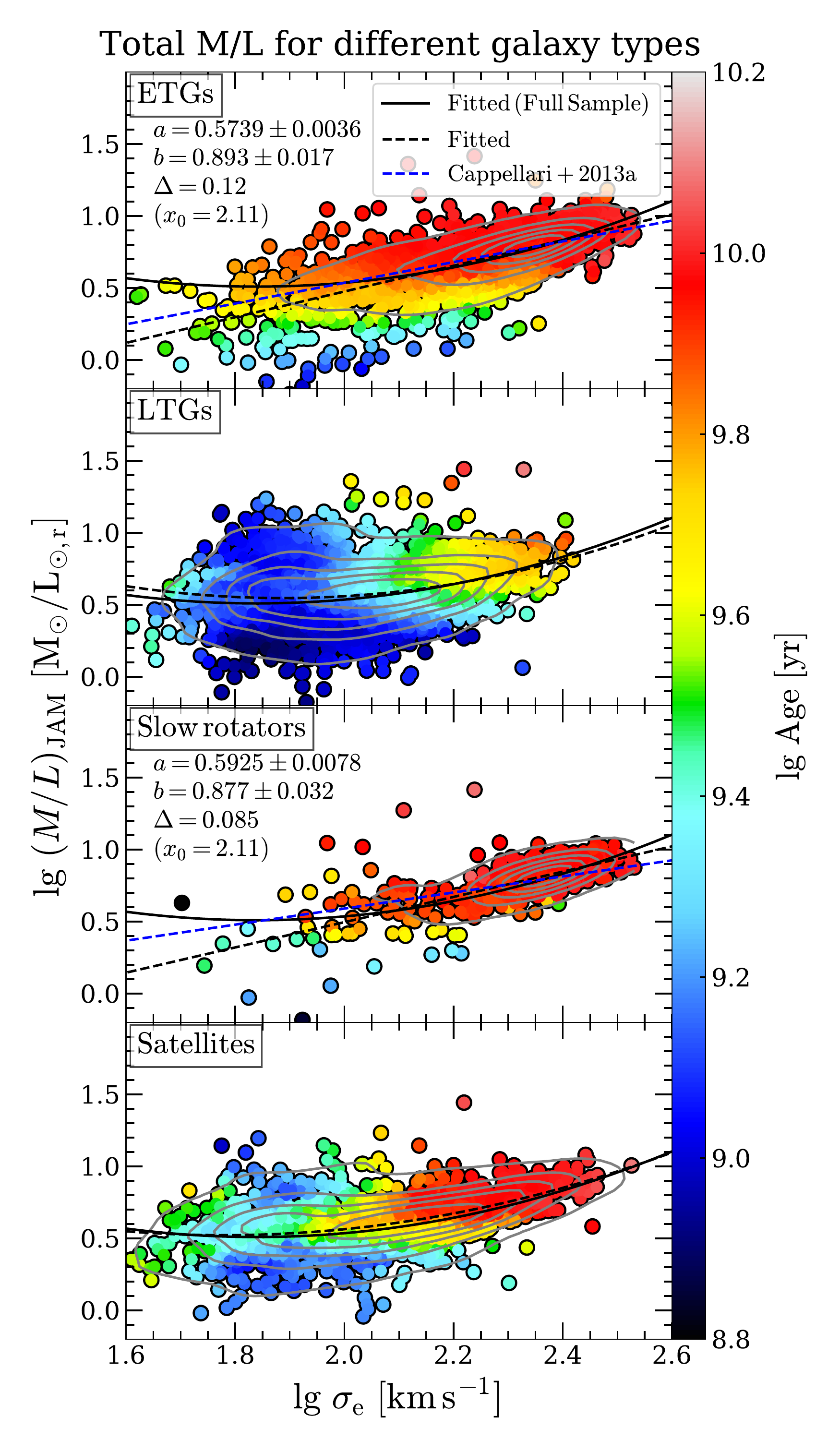}
    \caption{The mass-follows-light models inferred dynamical mass-to-light ratios as a function of $\sigma_{\rm e}$ for ETGs, LTGs, slow rotators, and satellites (from top to bottom). The symbols, lines, curves, and grey contours are the same as \autoref{fig:MLjam_sigma_agetype}. The relations obtained from $\rm ATLAS^{3D}$ \citep{Cappellari2013a} for ETGs or slow rotators are shown as blue dashed lines.}
    \label{fig:MLjam_sigma_type}
\end{figure}

In the top two panels of \autoref{fig:MLjam_sigma_type}, we present the $(M/L)_{\rm JAM}-\sigma_{\rm e}$ relations for ETGs and LTGs. For each panel, the black solid line is the best-fitting relation derived from the full sample, while the black dashed line is the best-fitting relation for the corresponding subset of galaxies. For the ETGs, we perform a linear fitting using the {\sc lts\_linefit} procedure and obtain the best-fitting relation with a scatter of $\Delta=0.12$ dex (32 per cent) and a slope of $b=0.893$. The scatter is consistent with the ETGs in $\rm ATLAS^{3D}$ \citep[0.11dex or 29 per cent,][]{Cappellari2013a}, but larger than the scatters found in the ETGs of the Virgo cluster \citep[0.054 dex or 13 per cent,][]{Cappellari2013a} and the Coma cluster \citep[0.070 dex or 17 per cent,][]{Shetty2020}. The reduction in the scatter for the cluster member galaxies is due to the much smaller uncertainty in relative distance measurements between the galaxies. The slope ($b=0.893$) is slightly steeper than those found in $\rm ATLAS^{3D}$ ($b=0.72$) and the Coma cluster ($b=0.69$), which is likely caused by the sample selection bias: the MaNGA ETGs sample contains more massive galaxies and the curvature of the $(M/L)_{\rm JAM}-\sigma_{\rm e}$ relation clearly shows that the slope is steeper for more massive (or higher-$\sigma_{\rm e}$) galaxies. We also find that the parabolic relations are quite similar between the full sample and the LTGs (second panel of \autoref{fig:MLjam_sigma_type}).

In the third panel, we show the $(M/L)_{\rm JAM}-\sigma_{\rm e}$ relations for the slow rotators. The best-fitting straight line for the slow rotators is similar to the one of ETGs but tends to have a slightly larger intercept ($a=0.5925$) and steeper slope ($b=0.877$). Given that 96 per cent slow rotators are ETGs (see \autoref{sec:morph_env_lambda}), we conclude that the fast rotating ETGs have a slightly smaller $(M/L)$ than the slow rotators (or slow rotating ETGs), which agrees with the trend found in $\rm ATLAS^{3D}$ \citep[fig.~15 in][]{Cappellari2013a}. The direct comparison with the relation of $\rm ATLAS^{3D}$ slow rotators (blue dashed line in the third panel) also indicates the effect of sample selection: the MaNGA slow rotators tend to have higher $\sigma_{\rm e}$, thus the slope of the best-fitting relation is steeper. 

Our parabolic $(M/L)_{\rm JAM}-\sigma_{\rm e}$ relation is consistent with early indications of a qualitatively nonlinear trend by \citet[fig.~9]{Zaritsky2006} and \citet{Aquino-Ortiz2020}. However, the previous result was based on dynamical masses derived assuming galaxies follow a manifold, while ours are high-quality direct quantitative measurements from dynamical models of thousands of galaxies.

Previous studies had shown the minor effect of environment on the $(M/L)-\sigma_{\rm e}$ for ETGs \citep{Cappellari2006,vanderMarel2007,Shetty2020}. We confirm this finding and extend it to LTGs from the bottom panel of \autoref{fig:MLjam_sigma_type}, in which the best-fitting relation for satellite galaxies is nearly identical to the one derived from the full sample. However, we also find some weak features of environmental effect for satellite galaxies: (i) the stellar ages of satellite galaxies are slightly older at fixed $\sigma_{\rm e}$; (ii) the scatter of $(M/L)_{\rm JAM}$ for satellite galaxies is smaller at $1.8<\lg (\sigma_{\rm e}/{\rm km\,s^{-1}})<2.0$, which is induced by the lack of very young satellite galaxies. The differences in stellar age demonstrate the picture: the satellites fall into the more massive dark halos and lose their gas under tidal stripping or ram pressure stripping, then the star formation ceases and the galaxies become quenched, finally the satellites are older than the central counterparts.

\subsection{The total-density slope vs. dispersion \texorpdfstring{$\overline{\gamma_{_{\rm T}}} - \sigma_{\rm e}$}{} relation}
\begin{figure}
    \centering
    \includegraphics[width=\columnwidth]{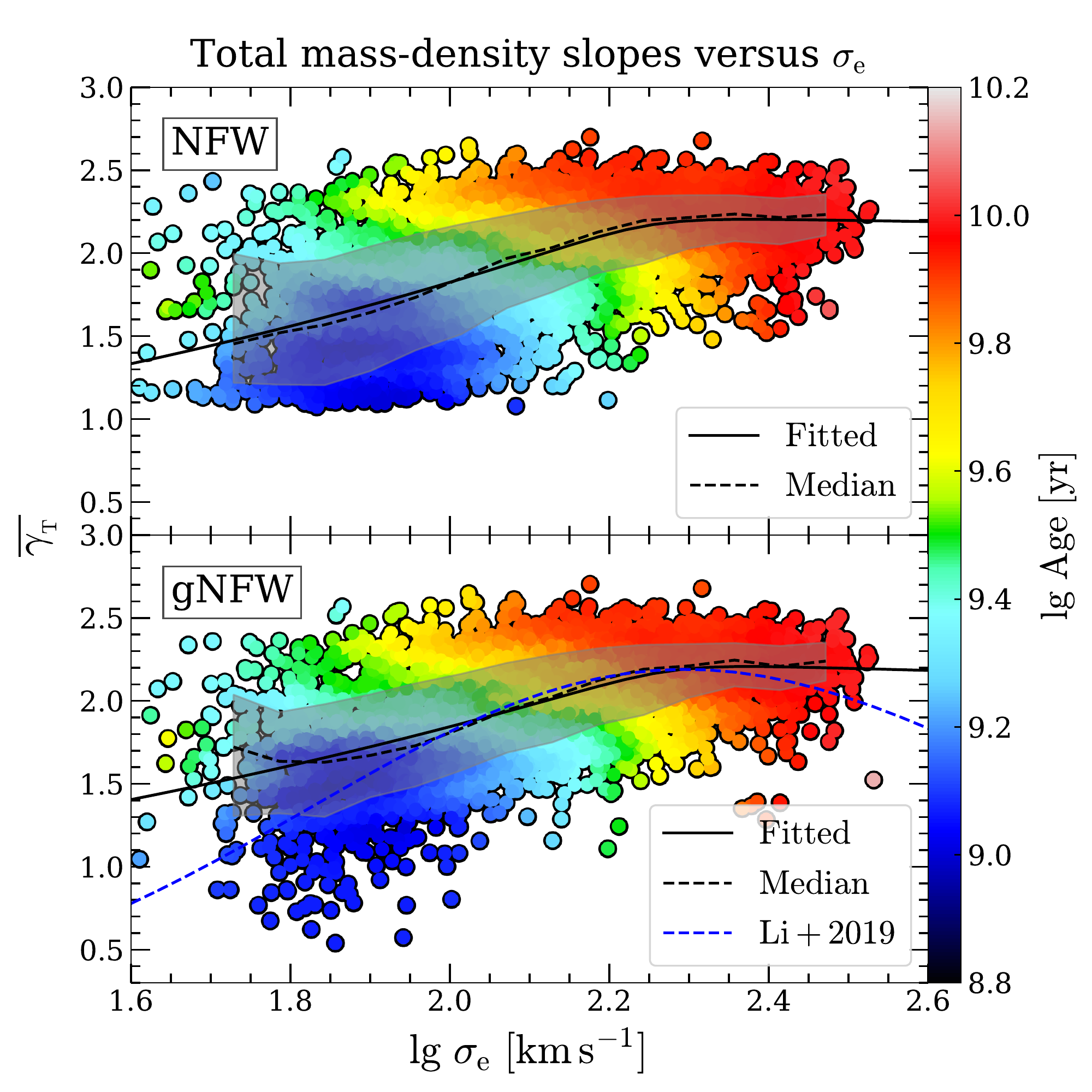}
    \caption{The mass-weighted total density slopes as a function of $\sigma_{\rm e}$ for different assumptions on dark matter halo. The symbols are coloured by stellar age. In each panel, the black dashed line represents the median value, while the grey shaded region denotes the [16th, 84th] percentile of values. The black solid lines are the best-fitting double power-law relations in the form of \autoref{eq:gammat_fit}, with $[A_{0}, \sigma_{\rm b}, \alpha, \beta, \gamma] = [2.17, 177, 11.03, -0.01, 0.34]$ for the NFW model (top) and $[2.18, 189, 11.13, -0.02, 0.30]$ for the gNFW model (bottom). The blue dashed curve is the best-fitting relation in \citet[eq.~12]{Liran2019}.}
    \label{fig:gammat_sigma}
\end{figure}

\autoref{fig:gammat_sigma} presents the relations of $\overline{\gamma_{_{\rm T}}}-\sigma_{\rm e}$, where the $\overline{\gamma_{_{\rm T}}}$ is the mass-weighted total density slope within $1 R_{\rm e}$ (see eq. 21 in \citetalias{Zhu2023a}), written as 
\begin{equation}\label{eq:density2}
    \overline{\gamma_{_{\rm T}}} \equiv \frac{1}{M_{\rm T}(<R_{\rm e})}\int_0^{R_{\rm e}}-\frac{\mathrm{d}\lg{\rho_{_{\rm T}}}}{\mathrm{d}\lg{r}} 4\pi r^2\rho_{_{\rm T}}(r)\mathrm{d}r = 3-\frac{4\pi R_{\rm e}^3\rho_{_{\rm T}}(R_{\rm e})}{M_{\rm T}(<R_{\rm e})}\,.
\end{equation}
The relations can be described using
\begin{equation}
	\label{eq:gammat_fit}
	\overline{\gamma_{_{\rm T}}} = A_{0}\left(\frac{\sigma_{\rm e}}{\sigma_{b}}\right)^{\gamma}\left[\frac{1}{2}+\frac{1}{2}\left(\frac{\sigma_{\rm e}}{\sigma_{b}}\right)^{\alpha}\right]^{\frac{\beta-\gamma}{\alpha}},
\end{equation}
with $[A_{0}, \sigma_b, \alpha, \beta, \gamma] = [2.17, 177, 11.03, -0.01, 0.34]$ for the NFW model and $[2.18, 189, 11.13, -0.02, 0.30]$ for the gNFW model. We find that $\overline{\gamma_{_{\rm T}}}$ decrease rapidly (the total slopes become steeper) with increasing $\sigma_{\rm e}$ at $\sigma_{\rm e}<\sigma_{b}$, with power slope $\gamma\approx0.30$ while at higher $\sigma_{\rm e}$ the relation becomes essentially constant (power slope $\beta\approx0$) with mean value $\overline{\gamma_{_{\rm T}}}\approx2.2$.

The constancy and value of the total slope above $\sigma_{\rm b}$ accurately agrees with the originally reported "universal" slope $\overline{\gamma_{_{\rm T}}}\approx2.2$ for ETGs out to 4$R_{\rm e}$ \citep{Cappellari2015,Serra2016,Bellstedt2018}. Extending the trend for lower $\sigma_{\rm e}$ and using data with more limited spatial extent, \citet{Poci2017} noted that there was a break in the $\overline{\gamma_{_{\rm T}}}-\sigma_{\rm e}$ relation for ETGs around $\lg\sigma_{\rm e}\la2.1$ and below that value the profiles were becoming more shallow. Both the nearly-constant region and the turnover were seen much more clearly by \citet{Liran2019}, using both spirals and ETGs from MaNGA, as we do here, but on a smaller sample of galaxies. Our results confirm and strengthen all previous trends on the $\overline{\gamma_{_{\rm T}}}-\sigma_{\rm e}$ relation, although the trend at low-$\sigma_{\rm e}$ is slightly different from the one in \citet{Liran2019} due to the updated stellar kinematics of MaNGA DAP \citep{Law2021}. As concluded in \citet{Law2021}, the scientific results based on the velocity dispersion far below the instrumental resolution (70 $\rm km\,s^{-1}$) should be reevaluated, leading to the higher $\sigma_{\rm e}$ and steeper total slopes of the final MaNGA data release at the low-$\sigma_{\rm e}$ end when compared to \citet{Liran2019}. The scatter decreases from $0.37$ to $0.12$ (a median value of $0.27$) with increasing $\sigma_{\rm e}$.

\begin{figure}
    \centering
    \includegraphics[width=\columnwidth]{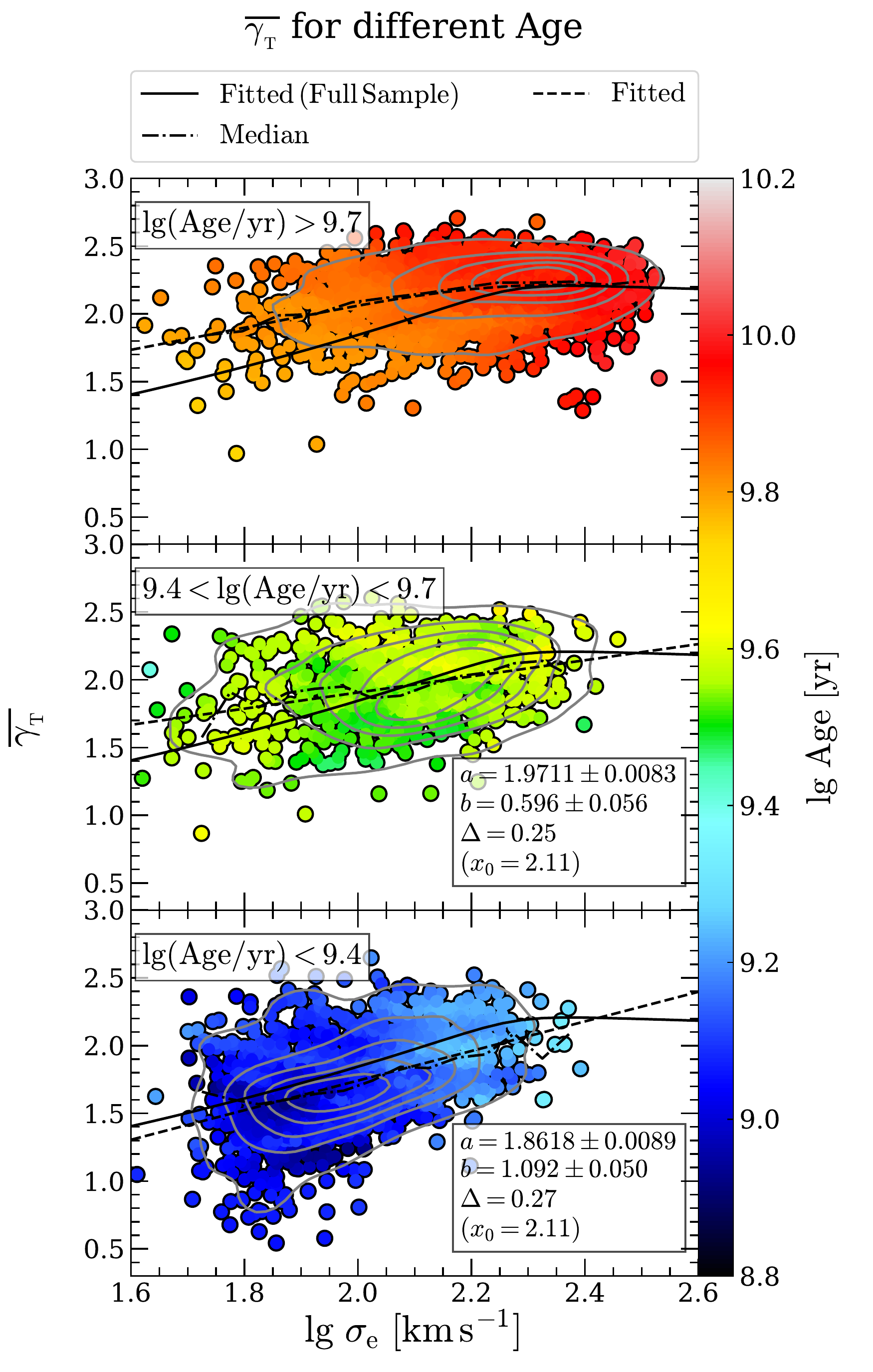}
    \caption{The gNFW models inferred mass-weighted total density slope as a function of $\sigma_{\rm e}$ for galaxies with different stellar ages. From top to bottom, the relations for old galaxies, intermediate galaxies, and young galaxies are presented. The grey contours are the kernel density estimate for the galaxy distribution. The symbols are coloured by stellar age. The black solid curve is the best-fitting relation derived from the full sample, while the black dashed curves(lines) represent the best-fitting relations of the corresponding subsample (the best-fitting parameters are shown in \autoref{tab:relations}).}
    \label{fig:gammat_sigma_agetype}
\end{figure}

Here we also look at the dependency of $\overline{\gamma_{_{\rm T}}}$ on the age of the stellar population. As shown in \autoref{fig:gammat_sigma}, we find that the $\overline{\gamma_{_{\rm T}}}$ varies with stellar age at fixed $\sigma_{\rm e}$, indicating the correlations between total density slopes and stellar age. This is consistent with the difference in total slopes of ETGs and LTGs reported by \citet{Liran2019} and with the difference in total slopes between young/old galaxies at fixed $\sigma_{\rm e}$ described by \citet{Lu2020}. In \autoref{fig:gammat_sigma_agetype}, we present the $\overline{\gamma_{_{\rm T}}}-\sigma_{\rm e}$ relations for galaxies with different age. For the old galaxies (the top panel in \autoref{fig:gammat_sigma_agetype}), a turnover of the relation is found, and the turnover point, $(\sigma_{\rm e}\approx179\,{\rm km\,s^{-1}})$, is slightly small than the one for the full sample. For the galaxies with younger stellar population (the second panel in \autoref{fig:gammat_sigma_agetype}), the relation monotonically increases with increasing $\sigma_{\rm e}$ with a slope of $b=0.596$. A similar monotonically increasing $\overline{\gamma_{_{\rm T}}}-\sigma_{\rm e}$ relation but with a steeper slope ($b=1.092$) is found for the youngest galaxies (the third panel of \autoref{fig:gammat_sigma_agetype}).

\begin{figure}
    \centering
    \includegraphics[width=\columnwidth]{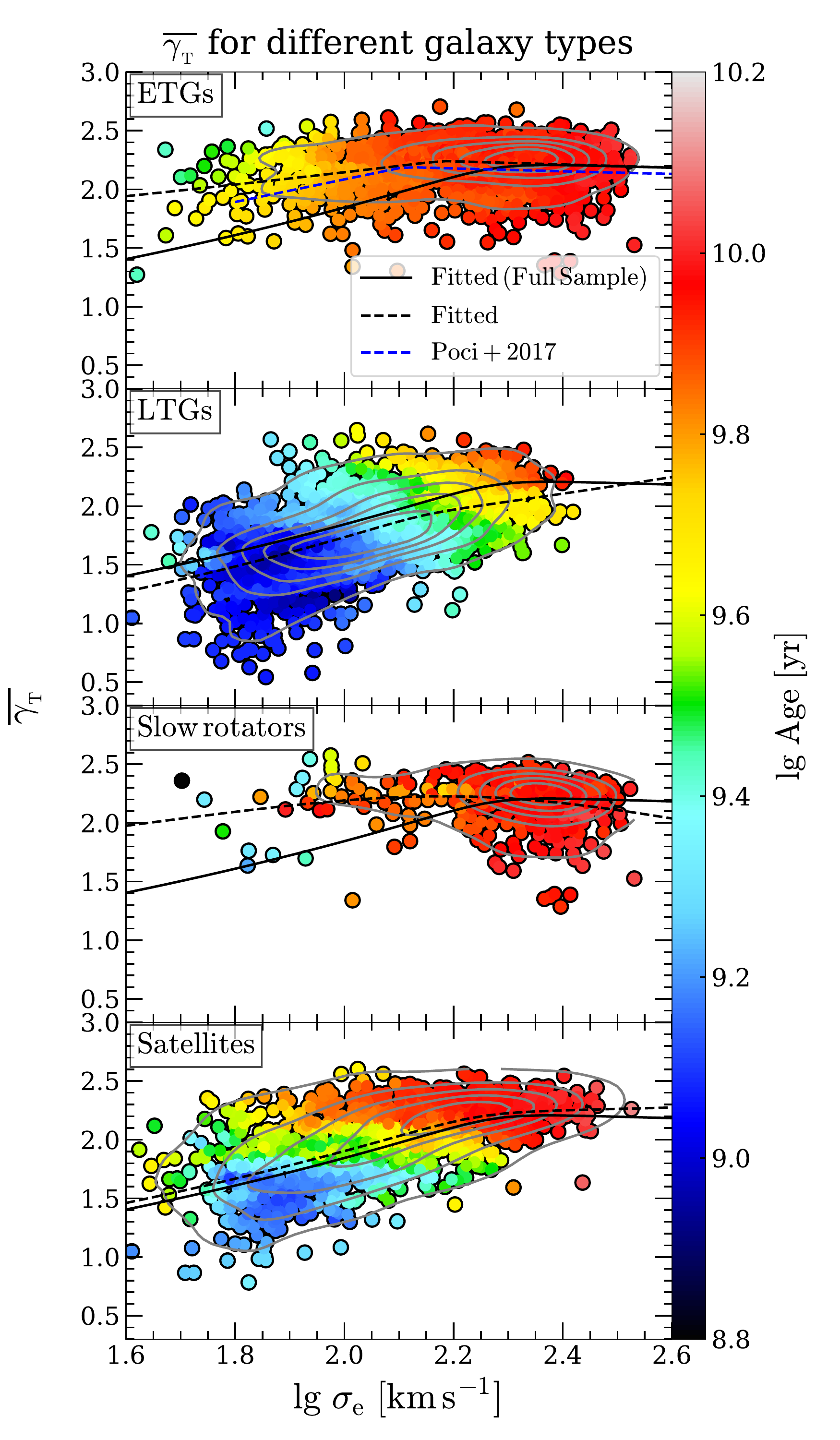}
    \caption{The gNFW models inferred mass-weighted total density slope as a function of $\sigma_{\rm e}$ for ETGs, LTGs, slow rotators, and satellites (from top to bottom). The symbols, curves, and grey contours are the same as \autoref{fig:gammat_sigma_agetype}. The relation obtained from $\rm ATLAS^{3D}$ plus SLACS \citep{Poci2017} is shown as the blue dashed line in the top panel.}
\label{fig:gammat_sigma_type}
\end{figure}

In \autoref{fig:gammat_sigma_type}, we present the relations for the ETGs (top panel), LTGs (second panel), slow rotators (third panel), and satellite galaxies (bottom panel). For the galaxies with different morphology (i.e. ETGs and LTGs), we find that the total slopes of LTGs are shallower than those of ETGs. This is consistent with the finding in \autoref{fig:gammat_sigma_agetype} that the galaxies with younger stellar age have shallower total density slopes. Specifically, the $\overline{\gamma_{_{\rm T}}}-\sigma_{\rm e}$ relation of MaNGA ETGs qualitatively agrees with that of $\rm ATLAS^{3D}$ plus SLACS \citep{Poci2017}. Compared to the relation derived from the full sample, the total slopes of slow rotators are shallower in the range of $\lg(\sigma_{\rm e}/{\rm km\,s^{-1}})<2.25$ (consistent with the trend of ETGs). The trend of satellite galaxies is similar to that of the full sample (dominated by the central galaxies) but is systematically steeper by $\approx 0.1$, which had been found in \citet{Liran2019}. We only show the empirical relations in this section, the more detailed study on the total density slopes and the comparison with the predictions of cosmological simulations is presented in \citet{Li2023}.

\subsection{The \texorpdfstring{$f_{\rm DM}(<R_{\rm e}) - M_{\ast}$}{} relation}
\begin{figure*}
    \centering
    \includegraphics[width=2\columnwidth]{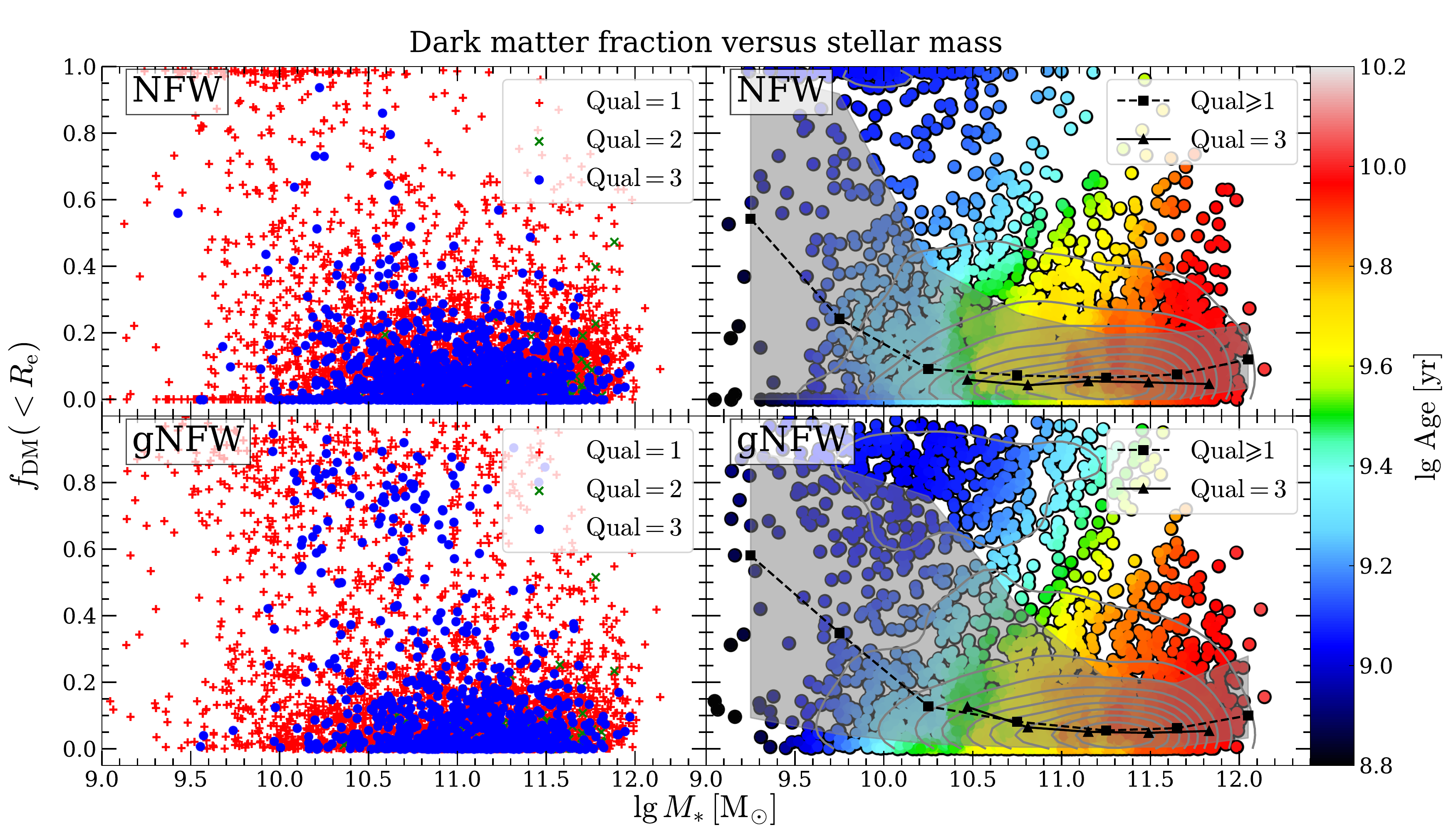}
    \caption{The dark matter fractions within a sphere of $R_{\rm e}$ as a function of Salpeter IMF-based stellar mass $M_{\ast}$, which is taken from the SPS models (\autoref{eq:MsSPS}). From top to bottom, the results of NFW and gNFW models are shown,  with symbols coloured by different modelling qualities and stellar age in the left and right panels, respectively. In the right panels, the dashed curves represent the median values in different mass bins for the $\rm Qual\geqslant1$ galaxies, while the gray shaded region is enclosed by [16th, 84th] percentile values. The solid curve is the median relation for $\rm Qual=3$ galaxies. The grey contours are the kernel density estimate for the galaxy distribution.}
    \label{fig:fdm_Ms_qual}
\end{figure*}

As shown in fig.~11 of \citetalias{Zhu2023a}, there is no systematic bias in dark matter fraction $f_{\rm DM}(<R_{\rm e})$ between different assumptions on the orientation of velocity ellipsoid (i.e. JAM$_{\rm cyl}$ vs. JAM$_{\rm sph}$). However, the mass models with different assumptions on the dark matter component may significantly affect the dark matter fraction for a small subset of galaxies (fig.~14 in \citetalias{Zhu2023a}), thus we use two mass models (NFW and gNFW models) to investigate the robustness of $f_{\rm DM}(<R_{\rm e})-M_{\ast}$ relations. Furthermore, we also select the galaxies with $|f_{\rm DM, cyl}-f_{\rm DM, sph}|<0.1$ to avoid the possible effect of bad modelling, as suggested in tab.~2 of \citetalias{Zhu2023a}. In the left panels of \autoref{fig:fdm_Ms_qual}, the dark matter fraction for both models are presented (NFW model in the top panel, gNFW model in the bottom panel), with coloured symbols corresponding to different modelling qualities. In agreement with previous studies \citep{Cappellari2013a,Shetty2020}, the galaxies with the best modelling quality statistically have lower dark matter fractions. This is most likely due to the dark matter estimates being unreliable for low-quality data. For this reason, we will show the dark matter fraction relations for different modelling qualities in the following discussions.

In the top right panel of \autoref{fig:fdm_Ms_qual}, we find a trend of $f_{\rm DM}(<R_{\rm e})-M_{\ast}$ relation for $\rm Qual\geqslant1$ galaxies: the median dark matter fraction rapidly decreases with increasing stellar mass within the range of $M_{\ast}<10^{10}{\rm M_{\odot}}$ (from 40 per cent to 10 per cent), and remains nearly unchanged for $M_{\ast}>10^{10}{\rm M_{\odot}}$ galaxies ($\approx10$ per cent). For the galaxies with better modelling quality (i.e. $\rm Qual=3$), the trend is quantitatively unchanged. The scatter, which is defined as (84th percentile - 16th percentile)/2, also decreases with increasing stellar mass from 50 per cent to 10 per cent. A similar trend is also found for the gNFW model (bottom right panel of \autoref{fig:fdm_Ms_qual}), thus the trend of $f_{\rm DM}(<R_{\rm e})-M_{\ast}$ is not affected by mass model differences. 

\begin{figure}
    \centering
    \includegraphics[width=\columnwidth]{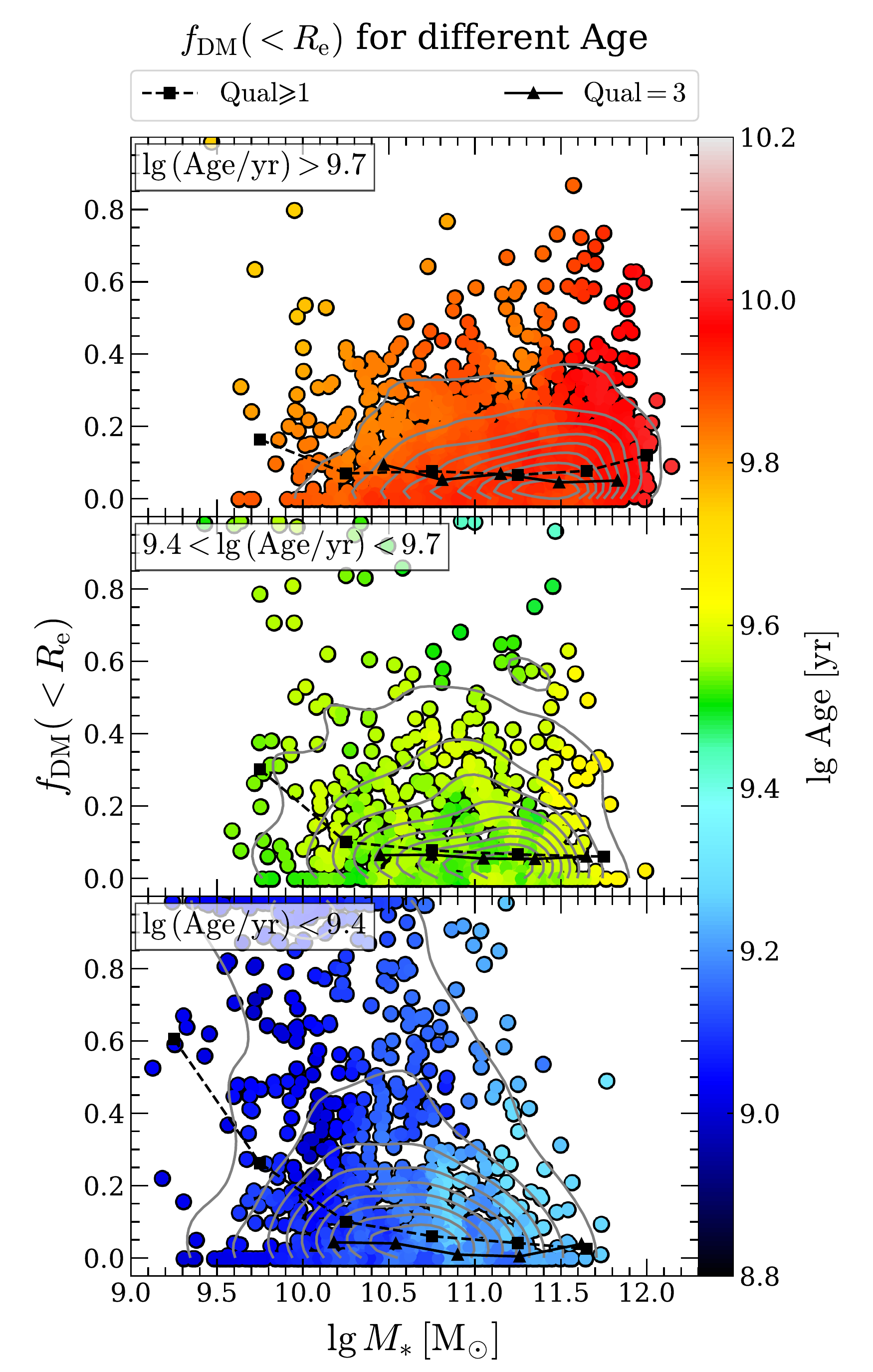}
    \caption{The NFW models inferred dark matter fractions within a sphere of $R_{\rm e}$ as a function of Salpeter IMF-based $M_{\ast}$ for old galaxies, intermediate galaxies, and young galaxies (from top to bottom). The symbols, curves, and grey contours are the same as \autoref{fig:fdm_Ms_qual}.}
    \label{fig:fdm_Ms_agetype}
\end{figure}

Moreover, we find that the dark matter fractions of older galaxies are lower at fixed stellar mass, indicating the diverse dark matter fraction for different stellar ages (the top panel of \autoref{fig:fdm_Ms_agetype}). In \autoref{fig:fdm_Ms_agetype}, the generally low dark matter fractions (with a median of 7 per cent and 90th percentile of 25 per cent for $\rm Qual\geqslant1$ galaxies) are found for the galaxies with $\rm \lg (Age/yr)>9.7$. The younger $\rm Qual\geqslant1$ galaxies with $\rm 9.4<\lg (Age/yr)<9.7$ also have $M_{\ast}$-independent and low dark matter fractions with a median of 8 per cent and 90th percentile of 38 per cent. We find more $\rm Qual\geqslant1$ galaxies with high dark matter fraction (90th percentile of 80 per cent) in the stellar age bin of $\rm \lg (Age/yr)<9.4$, although the median value (9 per cent) still indicates that this sample is dominated by galaxies with low dark matter fraction. The conclusions do not change if we only account for the galaxies with the best modelling quality ($\rm Qual=3$), as the relations are nearly identical in the range of $10^{10}{\rm M_{\odot}}<M_{\ast}<10^{11.5}{\rm M_{\odot}}$.

\begin{figure}
    \centering
    \includegraphics[width=\columnwidth]{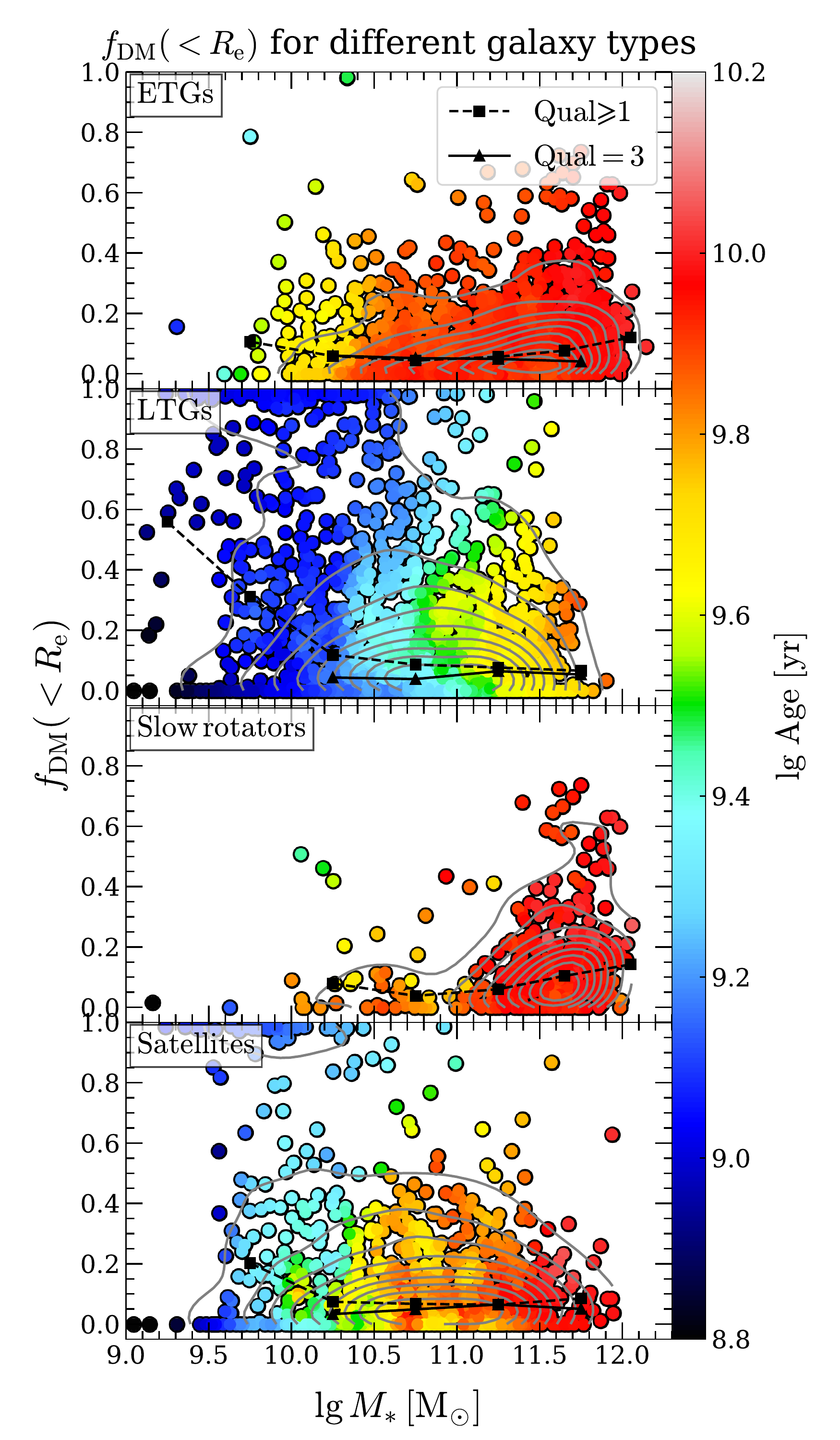}
    \caption{The NFW models inferred dark matter fractions within a sphere of $R_{\rm e}$ as a function of Salpeter IMF-based $M_{\ast}$ for ETGs, LTGs, slow rotators, and satellites (from top to bottom). The symbols, curves, and grey contours are the same as \autoref{fig:fdm_Ms_qual}.}
    \label{fig:fdm_Ms_type}
\end{figure}

To explore the effects of galaxy types on dark matter fraction, we also present the relations for different subsamples (ETGs, LTGs, slow rotators, and satellite galaxies) in \autoref{fig:fdm_Ms_type}. The most significant difference in $f_{\rm DM}(<R_{\rm e})-M_{\ast}$ relations lies in the morphology of galaxies: the dark matter fractions of ETGs remain nearly constant for different $M_{\ast}$, while the LTGs' dark matter fractions strongly correlate with $M_{\ast}$. For ETGs, we find the generally low dark matter fractions, with a median value of $f_{\rm DM}(<R_{\rm e})=7$ per cent for the $\rm Qual\geqslant1$ sample (6 per cent for the $\rm Qual=3$ sample). In addition, 90 per cent of $\rm Qual\geqslant1$ ETGs have $f_{\rm DM}(<R_{\rm e})<23$ per cent, while the value becomes $f_{\rm DM}(<R_{\rm e})<15$ per cent for $\rm Qual=3$ ETGs. The results agree with recent studies based on detailed stellar dynamical models: \citet{Cappellari2013a} found a median $f_{\rm DM}(<R_{\rm e})=13$ per cent for the full sample of $\rm ATLAS^{3D}$ and $f_{\rm DM}(<R_{\rm e})=9$ per cent for the sample of best models (i.e. quality>1 in tab.~1 of \citealt{Cappellari2013a}); \citet{Posacki2015} constructed JAM models for 55 ETGs of the Sloan Lens ACS (SLACS) sample and found a median $f_{\rm DM}(<R_{\rm e})=14$ per cent; \citet{Shetty2020} investigated 148 ETGs in the Virgo cluster and reported a median value of $f_{\rm DM}(<R_{\rm e})=25$ per cent and 90th percentile value of 34.6 per cent. Our values also are broadly consistent with the $f_{\rm DM}(<R_{\rm e})$ range of earlier stellar dynamics studies \citep[e.g.][]{Gerhard2001,Cappellari2006,Thomas2007,Williams2009}, which are obtained from a much smaller sample, but on the lower limit. Our dark-matter fractions tend to be smaller than those derived for a subset of 161 SAMI passive galaxies by \cite{Santucci2022}. This may be due to the lower data quality and more general models used in that study. Other studies based on gravitational lensing (e.g. a median of 23 per cent in \citealt{Auger2010}) or the joint lensing/dynamics analysis (e.g. a median of 31 per cent with assumed Salpeter IMF in \citealt{Barnabe2011}) also support the low dark matter fraction in ETGs. 

As opposed to the invariant low dark matter fractions in ETGs, the LTGs have diverse dark matter fractions for different stellar masses. In the second panel of \autoref{fig:fdm_Ms_type}, the $f_{\rm DM}(<R_{\rm e})$ decrease with increasing stellar mass till $M_{\ast}=10^{10}{\rm M_{\odot}}$, above which a flattening trend is observed. A similar trend had been reported by \citet{Tortora2019}, which uses the HI rotation curves of 152 LTGs in the SPARC sample \citep{Lelli2016} to infer the dark matter fraction by assuming a constant $K$-band stellar mass-to-light ratio of $0.6\,{\rm M_{\odot}/L_{\odot, K}}$. \citet{Courteau2015} found a monotonically decreasing trend which differs from the one in MaNGA, but their result was determined from a much smaller sample and hence suffered from larger uncertainty. However, we can not use the sample of the best quality ($\rm Qual=3$) to confirm the rapidly decreasing trend in the range of $M_{\ast}<10^{10}{\rm M_{\odot}}$, due to too few $\rm Qual=3$ galaxies within that mass range.

In the third panel of \autoref{fig:fdm_Ms_type}, we present the relation for slow rotators, which are dominated by the ETGs with old stellar population. In consistent with the trends of old galaxies (\autoref{fig:fdm_Ms_agetype}) and ETGs (\autoref{fig:fdm_Ms_type}), the slow rotators have $M_{\ast}$-independent low dark matter fraction (with a median of 8 per cent). We also present the dark matter fractions of satellite galaxies in the fourth panel of \autoref{fig:fdm_Ms_type}. Compared to the full sample, the satellite galaxies are older at fixed $\sigma_{\rm e}$ (especially at low-$\sigma_{\rm e}$ end), which is due to the environmental effects on the quenching of satellites \citep{Peng2012,Wang2020}. The satellite galaxies with old stellar age tend to have low dark matter fractions (with a median of 7 per cent), in agreement with the trend found in \autoref{fig:fdm_Ms_agetype}.

\subsubsection{Effects of $M_{\ast}/L$ gradients}
\label{sec:fdm_MLgrad}

When deriving total density profiles or total $M/L$, our dynamical models are formally correct, regardless of possible gradients in the stellar $M/L$, as long as our parametrization of the total density is sufficiently flexible to describe the real one. This is because the models only need to know the distribution of the luminous tracer population, from which we derive the kinematics, which is well approximated by the observed surface brightness. The models do not need to know the composition of the total density. However, when we decompose the total density into luminous and dark components, our results obviously depend on our assumption for the stellar $M/L$. We know that the adopted assumption of spatially constant stellar mass-to-light ratio in our models is just an approximation, as the stellar population gradients (including age, metallicity, and stellar mass-to-light ratio) in MaNGA galaxies had been reported \citep[][]{Zhengzheng2017,Goddard2017,Lihongyu2018,Dominguez-Sanchez2019,Ge2021,Parikh2021,Lu2023a}. Compared to the assumption of spatially constant stellar mass-to-light ratio, the $M_{\ast}/L$ with negative radial gradients will steepen the stellar mass-density profile, while the positive gradients of $M_{\ast}/L$ have the opposite effect. Since the dynamical models only put direct constraints on the total mass distribution, the decomposition between luminous and dark matter components is based on the assumption of more extended dark matter mass distribution (i.e. the shallower mass-density slope), which allows us to constrain the contribution of dark matter. Thus the stellar mass or the dark matter fractions inferred from dynamical models can be potentially affected by the steeper (shallower) stellar mass-density slopes when accounting for the negative (positive) $M_{\ast}/L$ gradients \citep[e.g.][]{Bernardi2018}. 

As discussed in sec.~3.3.2 of \citetalias{Zhu2023a}, the total density profile still can be correctly estimated for the models with constant $M_{\ast}/L$. In order to study the effects of $M_{\ast}/L$ gradients, we introduce these gradients into the stellar mass-density profiles and rerun the decomposition between luminous/dark components. We measured the radial profiles using \textsc{mge\_radial\_mass} in \textsc{JamPy} for both the stars, after applying $M_{\ast}/L$ gradients, and DM. Then we use these profiles to perform a least-squares fitting to the total density derived in the same way from the original models with constant $M_{\ast}/L$. The fitting is performed within the region where the kinematic data is available. For a given $M_{\ast}/L$ of the innermost luminous Gaussian and the $M_{\ast}/L$ gradient, we calculate the expected $M_{\ast}/L$ values at all luminous Gaussians' dispersions and assign them to the Gaussians. We use the $(M_{\ast}/L)_{\rm SPS}$ gradients in this test, which are taken from the stellar population analysis for the full sample of MaNGA galaxies (see \autoref{sec:sps} and \citetalias{Lu2023a} for more details). The $(M_{\ast}/L)_{\rm SPS}$ gradients are estimated within 1 $R_{\rm e}$ but we apply the gradients to the region where we have kinematic data. However, this gradient scaling which extends beyond 1 $R_{\rm e}$ will not affect the quantities measured within 1 $R_{\rm e}$, e.g. $f_{\rm DM}(<R_{\rm e})$, in this test. Note that the $(M_{\ast}/L)_{\rm SPS}$ are derived by assuming spatially constant Salpeter IMF \citep{Salpeter1955} and we don't account for the radial variation of IMF \citep[e.g.][]{vanDokkum2017} here \citep[see][for a review on the IMF variations]{Smith2020}.

\begin{figure*}
    \centering
    \includegraphics[width=2\columnwidth]{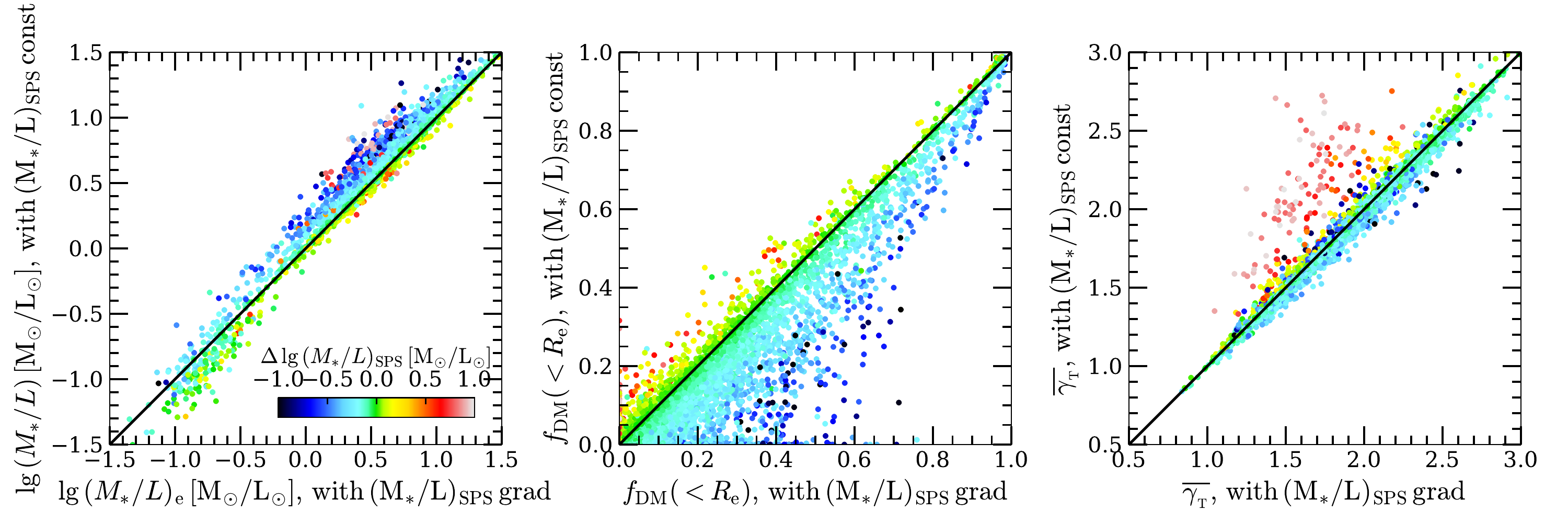}
    \caption{The comparison of the quantities ($M_{\ast}/L$, $f_{\rm DM}(<R_{\rm e})$, $\overline{\gamma_{_{\rm T}}}$) between the models with assumed spatially constant $M_{\ast}/L$ (y-axis) and the models with $M_{\ast}/L$ gradients (x-axis). The symbols are coloured by the $(M_{\ast}/L)_{\rm SPS}$ gradients, which are derived from the SPS models with assumed spatially constant Salpeter IMF (see \autoref{sec:sps} and \citetalias{Lu2023a} for more details about the gradients). At each panel, the black solid line represents the one-to-one relation.}
    \label{fig:MLgrad_MLconst}
\end{figure*}

\begin{figure*}
    \centering
    \includegraphics[width=2\columnwidth]{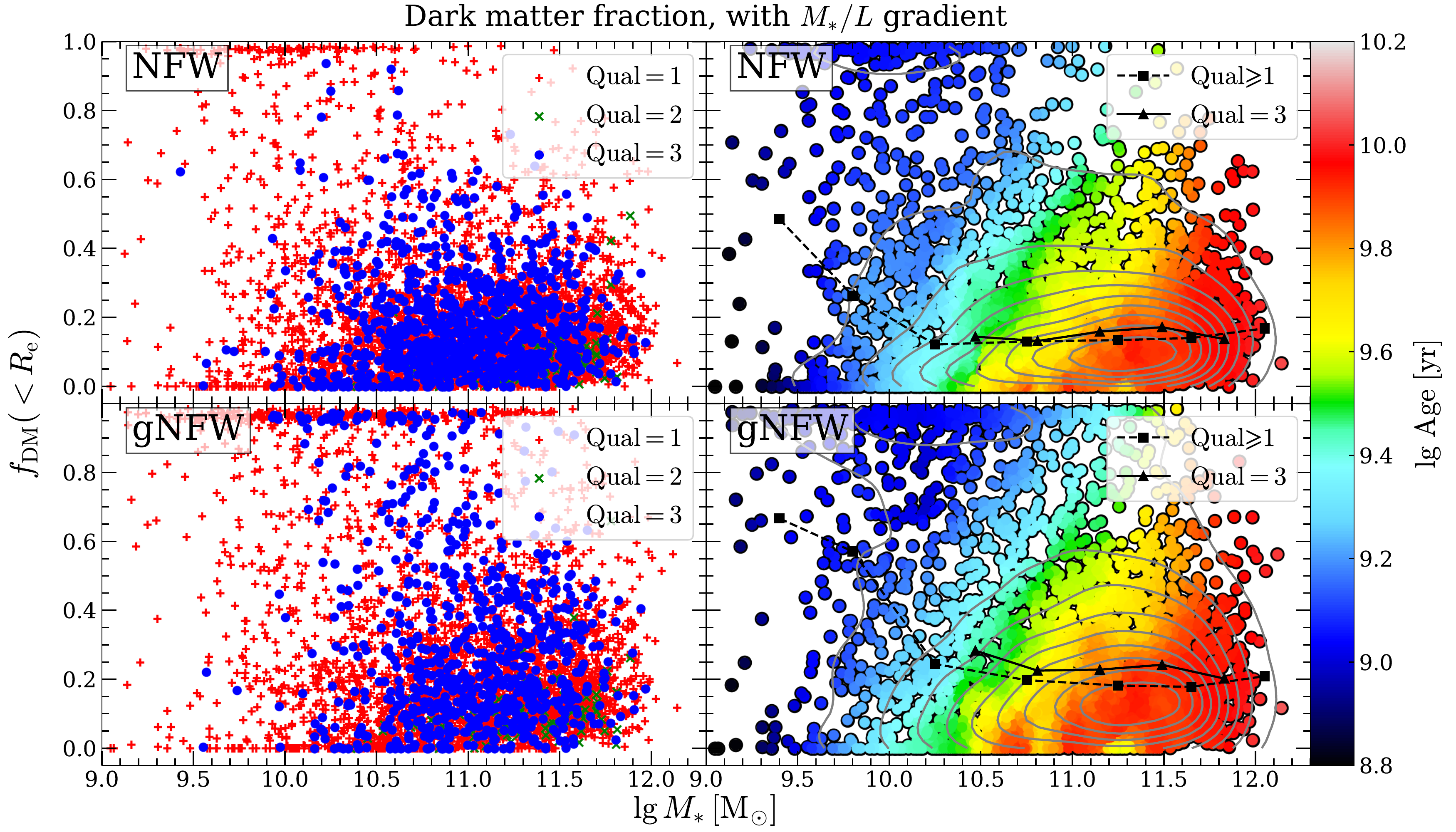}
    \caption{The dark matter fractions when incorporating the stellar mass-to-light ratio gradients into the stellar mass-density profile. The panels are the same as \autoref{fig:fdm_Ms_qual}.}
    \label{fig:fdm_Ms_MLgrad}
\end{figure*}

\autoref{fig:MLgrad_MLconst} shows the quantities derived from the models with $M_{\ast}/L$ gradients and those derived from the models with constant $M_{\ast}/L$. We use the \textsc{lts\_linefit} procedure to compare the two sets of quantities that are related to the decomposition in terms of luminous and dark components, i.e. the effective stellar mass-to-light ratios within $R_{\rm e}$ $(M_{\ast}/L)_{\rm e}$, the dark matter fractions within $R_{\rm e}$ $f_{\rm DM}(<R_{\rm e})$, and the total density slopes $\overline{\gamma_{_{\rm T}}}$. Given that $\sim$ 80 per cent of our sample ($\rm Qual\geqslant1$) are dominated by the galaxies with negative $M_{\ast}/L$ gradients and the median $(M_{\ast}/L)_{\rm SPS}$ gradient is -0.095 dex/$R_{\rm e}$, the dark matter fractions systematically but slightly increase. Finally, we find that the total density slopes $\overline{\gamma_{_{\rm T}}}$ are quite consistent for the two kinds of models, confirming the good fitting qualities of the models which incorporate the $M_{\ast}/L$ gradients. \autoref{fig:fdm_Ms_MLgrad} shows the $f_{\rm DM}(<R_{\rm e})-M_{\ast}$ relations using the models (including NFW and gNFW models) with $M_{\ast}/L$ gradients. Compared to the relations in \autoref{fig:fdm_Ms_qual}, the trend of rapidly decreasing of $f_{\rm DM}(<R_{\rm e})$ with increasing $M_{\ast}$ at the low-mass end and nearly unchanged $f_{\rm DM}(<R_{\rm e})$ at the high-mass end still exists, but the $f_{\rm DM}(<R_{\rm e})$ are systematically larger by $\sim$ 7 per cent for the NFW models ($\sim$ 13 per cent for the gNFW models).


\subsection{Dynamical properties on the mass-size plane}
As shown in \autoref{sec:FP_MP}, the MP, which consists of the mass $M_{\rm JAM}$, the velocity dispersion $\sigma_{\rm e}$, and the size $R_{\rm e}^{\rm maj}$, satisfies the scalar virial theorem very well especially when accounting for the inclination effects (the bottom panels of \autoref{fig:FP}). However, the edge-on view of the MP is thin and the galaxy properties smoothly vary with $\sigma_{\rm e}$, hence the tight MP does not contain too much useful information on the galaxy formation and evolution. For the face-on view of the MP, the galaxies with different properties located in different regions had been found using the SDSS single fibre spectrum \citep{Graves2009,Graves2010}. With the advent of spatially resolved spectroscopic observations (e.g. $\rm ATLAS^{3D}$, SAMI, MaNGA, LEGA-C) and the more accurate dynamical mass measurements, the inhomogeneous distributions of galaxy properties on the mass-size plane are confirmed \citep{Cappellari2013b,McDermid2015,Scott2017,Lihongyu2018,Cappellari2022,Barone2022}. Most of the previous studies focus on the stellar population properties (e.g. the stellar age, metallicity) on the mass-size plane, but the distributions of dynamical properties (e.g. the stellar angular momentum, total density slopes) also put constraints on the evolutionary path of galaxies \citep[see sec.~4.3 of][for a review]{Cappellari2016ARAA}.

In \autoref{fig:mass_size_plane}, we present the stellar velocity dispersions $\sigma_{\rm e}$, the deprojected specific stellar angular momentum proxy $\lambda_{\rm R_{e}}$, the total density slopes $\overline{\gamma_{_{\rm T}}}$, the dark matter fractions within an effective radius $f_{\rm DM}(<R_{\rm e})$ on the mass-size plane. In the top left panel of \autoref{fig:mass_size_plane}, the observed $\sigma_{\rm e}$ follows the constant $\sigma$ lines, which are predicted from the scalar virial equation $M_{\rm JAM}\equiv5\times R_{\rm e}^{\rm maj}\sigma^2/G$ \citep{Cappellari2006}, where $M_{\rm JAM}$ is derived from JAM models (\autoref{eq:Mjam}) and $R_{\rm e}^{\rm maj}$ is the semi-major axis of half-light elliptical isophote (\autoref{sec:jam}). In the top right panel, we show the deprojected $\lambda_{\rm R_e}$, which is approximately estimated from the observed one (given by \autoref{eq:lambdaRe}) by deprojecting the observed velocity to the edge-on view using the best-fitting inclination derived from JAM models. The deprojected $\lambda_{\rm R_e}$ do not quite follow lines of constant $\sigma_{\rm e}$. Instead, as previously found \citep[fig.~8]{Cappellari2013b}, we confirm that $\lambda_{\rm R_{\rm e}}$ is mainly driven by the stellar mass rather than $\sigma_{\rm e}$, with most slow rotators (the galaxies with $\lambda_{\rm R_{\rm e}}\la0.2$ and red colour in the top-right panel of \autoref{fig:mass_size_plane}) being present above a characteristic mass $M_{\rm crit}\approx2\times10^{11}$ M$_\odot$ as found by a number of studies \citep{Emsellem2011,Cappellari2013b,Cappellari2013apjl,Veale2017,Graham2018}. See review by \citet{Cappellari2016ARAA}. Additionally, we find that above $M_{\rm crit}$ the slow rotators are concentrated at the largest $\sigma_{\rm e}\ga200$ \kms. We also find a smaller decrease of $\lambda_{\rm R_{\rm e}}$ with $\sigma_{\rm e}$ and especially near the magenta curve above its break in \autoref{fig:mass_size_plane}, i.e. the zone of exclusion (ZOE) which is roughly described as $R_{\rm e}^{\rm maj}\propto M_{\rm JAM}^{0.75}$ above the break \citep{Cappellari2013b}. A similar trend of decreasing angular momentum being associated with larger bulges and galaxies deviating from the star-forming main sequence was discussed in \citet{Wang2020}. This trend, as well as the trends of stellar populations (see fig.~8 of \citetalias{Lu2023a}), is consistent with the distribution of galaxy morphological types on the mass-size plane: the slopes of individual Hubble types parallel to the ZOE, with the ETGs locating closer to the ZOE and the LTGs being further away from the ZOE \citep[][fig.~9]{Bender1992,Burstein1997,Cappellari2013b}. The overall trend of $\lambda_{\rm R_e}$ on the $(M, R_{\rm e})$ plane can be understood as the combination of two effects: (i) larger bulges make $\lambda_{\rm R_e}$ lower and produce a weak trend of decreasing $\lambda_{\rm R_e}$ with decreasing $R_{\rm e}$ at fixed stellar mass; (ii) massive slow rotators, which are likely the results of early dry mergers \citep[e.g.][]{Bezanson2009,Naab2009,Cappellari2016ARAA} produce the bump of low $\lambda_{\rm R_e}$ galaxies above $M_{\rm crit}$ and for $\sigma_{\rm e}\ga200$ \kms.

The bottom left panel of \autoref{fig:mass_size_plane} also shows the parallel distributions of the total density slopes $\overline{\gamma_{_{\rm T}}}$ along the direction of the ZOE above the break $M_{\rm JAM}=2\times10^{10}{\rm M_{\odot}}$, with the total density slopes become steeper moving towards the ZOE. This trend is qualitatively similar but much stronger than the one previously seen by the ATLAS$^{\rm 3D}$ survey in ETGs \citep[fig.~22c]{Cappellari2016ARAA}. This trend of steeper $\overline{\gamma_{_{\rm T}}}$ with increasing $\sigma_{\rm e}$ agrees with the $\overline{\gamma_{_{\rm T}}}-\sigma_{\rm e}$ relation (\autoref{fig:gammat_sigma}) but the constant $\overline{\gamma_{_{\rm T}}}$ lines are not strictly following the constant $\sigma_{\rm e}$ lines. Instead, the slight tilt between the constant $\sigma_{\rm e}$ lines and the constant $\overline{\gamma_{_{\rm T}}}$ lines is consistent with the scatter of the $\overline{\gamma_{_{\rm T}}}-\sigma_{\rm e}$ relation: at fixed $\sigma_{\rm e}$ (or $M_{\rm JAM}$), the $\overline{\gamma_{_{\rm T}}}$ is steeper for the galaxies with old stellar ages and smaller sizes. Moreover, for the old galaxy populations (or ETGs) close to the ZOE, a transition of $\overline{\gamma_{_{\rm T}}}$ from slightly steeper than isothermal ($\overline{\gamma_{_{\rm T}}}\ga2.4$) to nearly isothermal ($\overline{\gamma_{_{\rm T}}}\approx2.2$) moving towards the upper right of the plane is also observed. At the largest $\sigma_{\rm e}$, the steepest total slopes are not found for the most massive galaxies, also consistently with \citet[fig.~22c]{Cappellari2016ARAA}. This can be interpreted as due to slow rotators dominating the largest masses. They tend to have more shallow density profiles than fast rotators of similar $\sigma_{\rm e}$ due to the lack of gas dissipation.

We present the dark matter fractions on the mass-size plane in the bottom right panel of \autoref{fig:mass_size_plane}. Most of the galaxies have low dark matter fractions, while the galaxies with low masses, young stellar ages, high $\lambda_{\rm R_e}$, and shallow total density slopes tend to have non-negligible dark matter fractions. Compared to the distributions of $\lambda_{\rm R_{e}}$ and $\overline{\gamma_{_{\rm T}}}$, we find similar parallel sequences of dark matter fractions but a transition of $f_{\rm DM}(<R_{\rm e})$ for the old galaxies (or ETGs) along the direction of the ZOE is not observed. It is the first time that such a clear trend in $f_{\rm DM}(<R_{\rm e})$ was observed. It is a robust result, nearly model-independent, that can be qualitatively understood even without the need for quantitative dark-matter decompositions: the reason is that total slopes $\overline{\gamma_{_{\rm T}}}$ are significantly lower (more shallow) than most of the stellar densities of the galaxies in that region. Only (i) a significant dominance of dark matter combined with (ii) shallow dark matter profiles, can produce such flat total densities. The increase of $f_{\rm DM}(<R_{\rm e})$ is also consistent with our comparison between the total $(M/L)_{\rm JAM}$ and the stellar $(M_*/L)_{\rm SPS}$ derived from stellar population in \citetalias{Lu2023a}. It appears to be the reason for the parabolic form of the $(M/L)_{\rm JAM}-\sigma_{\rm e}$ relation \citep{Lu2023b}.

\section{Discussion}
\label{sec:discuss}
\begin{figure*}
    \centering
    \includegraphics[width=2\columnwidth]{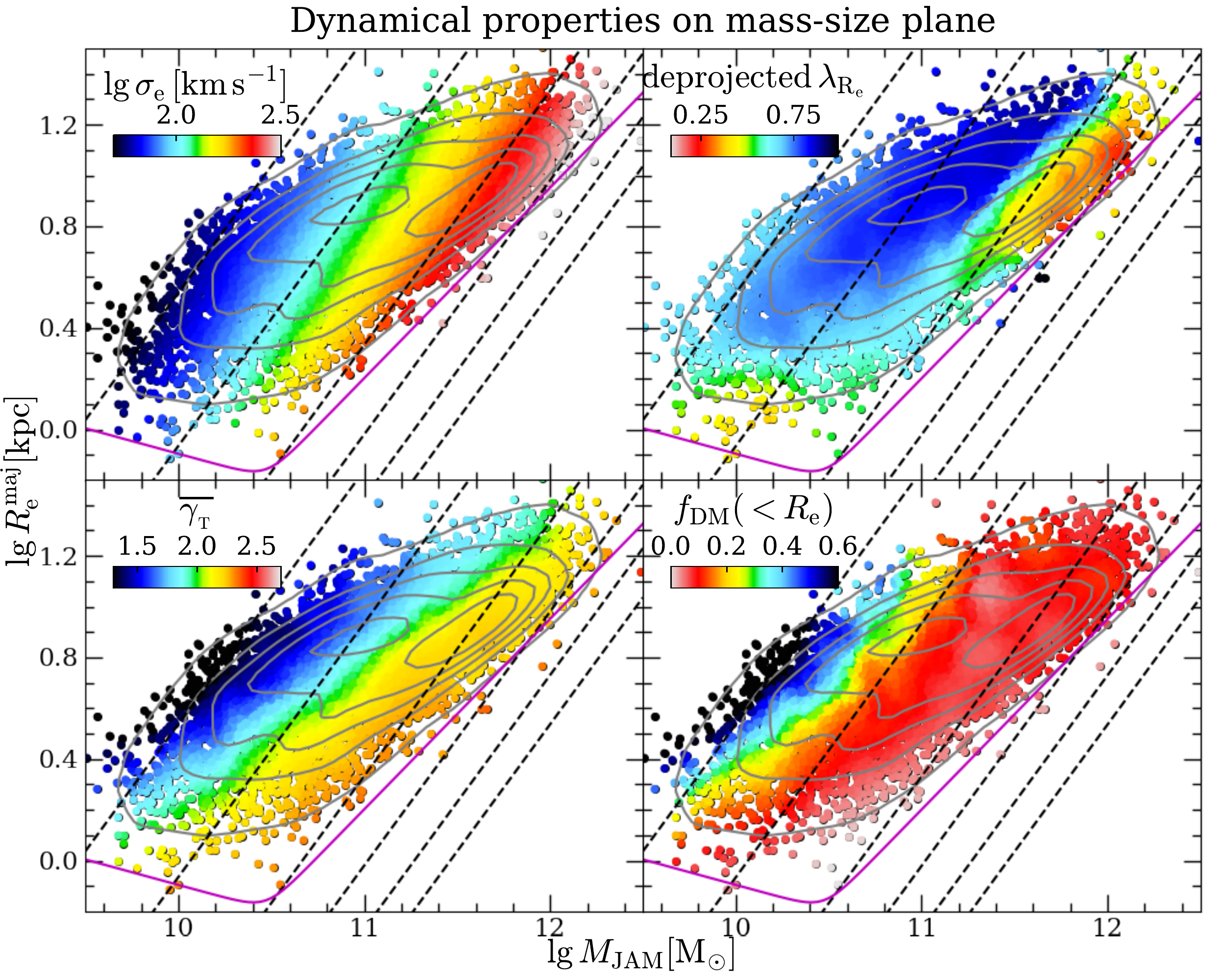}
    \caption{The distributions of dynamical properties ($\sigma_{\rm e}$, deprojected $\lambda_{\rm R_{e}}$, $\overline{\gamma_{_{\rm T}}}$, and $f_{\rm DM}(<R_{\rm e})$) on the $M_{\rm JAM}-R_{\rm e}^{\rm maj}$ plane. The deprojeced $\lambda_{\rm R_e}$ is obtained by deprojecting the $\lambda_{\rm R_e}$ to the edge-on view using the best-fitting inclination derived from JAM models. The distributions are smoothed by the \textsc{loess} software with \texttt{frac=0.05}. In each panel, the dashed lines correspond to the 50, 100, 200, 300, 400, and 500 $\rm km\,s^{-1}$ from left to right, which are calculated using the scalar virial equation $\sigma \equiv \sqrt{G M_{\rm JAM}/(5\times R_{\rm e}^{\rm maj})}$. The magenta curve shows the zone of exclusion (ZOE) defined in \citep{Cappellari2013b}, with the ZOE above $M_{\rm JAM}=2\times10^{10}{\rm M_{\odot}}$ is approximately $R_{\rm e}^{\rm maj}\propto M_{\rm JAM}^{0.75}$. The outliers beyond the ZOE are the galaxies with strong bars or face-on view, whose dynamical mass will be overestimated in JAM models \citep{Lablanche2012}. The grey contours show the kernel density estimate for the galaxy distribution.}
    \label{fig:mass_size_plane}
\end{figure*}

As suggested by the IFS results of nearby ETGs \citep{Cappellari2013b,Cappellari2016ARAA} and the observations of high-redshift ETGs \citep{vanderWel2008,vanDokkum2015,Derkenne2021}, two different build-up channels are needed to explain the evolutionary tracks on the mass-size plane: (i) the gas accretion or minor gas-rich mergers; (ii) the dry (i.e. gas-poor) mergers. Here, we revisit the two-phase evolution scenario in \autoref{fig:mass_size_plane}, as we use a larger sample that contains different morphological types. At the first stage when the LTGs (spirals) formed (i.e. the formation of stellar disks; \citealt{Mo1998}), the galaxies are still star-forming (young stellar age), have high stellar angular momentum (high $\lambda_{\rm R_{e}}$), and have small bulges (low $\sigma_{\rm e}$). The accreted cold gas falls in the inner regions of the LTGs, leading to enhanced in-situ star formation activities, steeper total density profiles \citep{Wangyunchong2019}, and lower central dark matter fractions. Meanwhile, the bulges grow (with increasing $\sigma_{\rm e}$ and decreasing $\lambda_{\rm R_{e}}$) and the star formation rates become lower (resulting in older stellar age) due to the bulge-related quenching mechanisms (e.g. the AGN feedback), until the galaxies become fully quenched \citep[e.g.][]{Chen2020}. During the in-situ star formation and subsequent quenching, both the galaxy masses and $\sigma_{\rm e}$ increase while the galaxy sizes (quantified by effective radius) decrease \citep{Cappellari2013b} or increase with a shallow slope of $R_{\rm e}\propto M_{\rm JAM}^{0.3}$ \citep{vanDokkum2015}. During the non-violent quenching mechanisms, the fast-rotating disk structures still remain but the stellar ages become older and the bulge fractions increase, leading to the transformation from LTGs to S0 galaxies (fast-rotating ETGs). 

As opposed to the decreasing \citep{Cappellari2013b} or slowly increasing \citep{vanDokkum2015} galaxy sizes in the gas accretion channel, the size evolution is more significant in dry mergers (including the major dry mergers that merge with comparably massive galaxy and the minor mergers that accrete many small satellites). For the major dry mergers, the galaxy sizes increase as the masses grow proportionally, with the $\sigma_{\rm e}$ remaining nearly unchanged, thus the galaxies move along the constant $\sigma_{\rm e}$ lines upwards. For the minor dry mergers, the sizes increase by a factor of four for doubling masses, while $\sigma_{\rm e}$ are twice smaller \citep{Bezanson2009,Naab2009}, leading to the evolutionary track that is steeper than the constant $\sigma_{\rm e}$ lines. The dry mergers reduce the stellar angular momentum through both the major mergers \citep{Hopkins2009,Zeng2021} and the minor ones \citep{Hopkins2009,Qu2010}, making it able to explain the transition of $\lambda_{\rm R_{e}}$ along the direction of the ZOE for the ETGs (top right panel in \autoref{fig:mass_size_plane}). Moreover, the evolution of total density slopes is also explainable: the slopes become slightly steeper than isothermal through the gas accretion process and then become shallower again through the dry mergers until reaching nearly isothermal \citep{Xudandan2017,Wangyunchong2019}. Given that there is little gas involved in the dry mergers, the dark matter fractions, as well as the stellar population properties (e.g. age, metallicity, stellar mass-to-light ratio) remain unchanged.

In summary, we find that the dynamical properties ($\sigma_{\rm e}$, $\lambda_{\rm R_{e}}$, $\overline{\gamma_{_{\rm T}}}$ and $f_{\rm DM}(<R_{\rm e})$) on the mass-size plane can be explained with the combination of two evolutionary channels: (i) gas accretion/gas-rich mergers; (ii) dry mergers. The young spirals grow their bulges via the enhanced central star formation induced by gas accretion, eventually leading to increasing stellar mass and $\sigma_{\rm e}$, steeper total density profiles, and lower central dark matter fractions, while the bulge-related quenching mechanisms (e.g. AGN feedback) tend to turn off the star formation until fully quenched. The non-violent quenching does not destroy the fast-rotating disks, thus the galaxies retain their high $\lambda_{\rm R_{e}}$. The gas accretion moves the galaxies from left to right on the mass-size plane while intersecting the constant $\sigma_{\rm e}$ lines with decreasing sizes \citep{Cappellari2013b} or mildly increasing sizes \citep{vanDokkum2015}. On the contrary, the dry mergers significantly increase the size \citep{Bezanson2009,Naab2009}, moving the galaxies upwards along the constant $\sigma_{\rm e}$ lines (major ones) or steeper (minor ones). Furthermore, the dry mergers lead to the slowing down of rotation \citep{Hopkins2009,Qu2010,Zeng2021}, the nearly isothermal total density profile \citep{Wangyunchong2019} and the nearly unchanged central dark matter fraction (due to little gas involved). The effect of dry mergers is more obvious for the ETGs close to the ZOE, of which the evolution is dominated by dry mergers.

\section{Summary}
\label{sec:summary}
In this paper, we present the dynamical scaling relations for $\sim$ 6000 nearby galaxies selected from the MaNGA SDSS-DR17 sample based on their dynamical modelling qualities (i.e. $\rm Qual\geqslant1$ as defined in \citetalias{Zhu2023a}). The dynamical quantities for the $\rm Qual\geqslant1$ galaxies had been demonstrated to have negligible systematic bias and small scatter between different models \citepalias[][tab.~3]{Zhu2023a}. Based on the dynamical quantities in \citetalias{Zhu2023a} and the stellar population properties in \citetalias{Lu2023a}, we investigate the fundamental plane (FP), the mass plane (MP), the total M/L, the total density slopes, the dark matter fractions, and the mass-size plane with combined dynamical and stellar population analysis. We classify the galaxies into subsamples based on their stellar ages: the old population ($\rm \lg\,(Age/yr)>9.7$), the intermediate population ($\rm 9.4<\lg\,(Age/yr)<9.7$), and the young population ($\rm \lg\,(Age/yr)>9.4$), and investigate how the relations change with stellar population. Moreover, we also present the relations for subsamples of different morphological types (ETGs or LTGs), satellites (classified by the Yang07 group catalogue), and slow rotators (occupied by ETGs). The dynamical scaling relations for the full sample and different subsamples are presented in \autoref{tab:relations}.

\begin{table*}
\centering
\caption{We summarize the empirical dynamical scaling relations (i.e. the FP, MP, $\lg\,(M/L)_{\rm JAM}-\lg\,\sigma_{\rm e}$, $\overline{\gamma_{_{\rm T}}}-\sigma_{\rm e}$, and the $f_{\rm DM}(<R_{\rm e})-\lg\,M_{\ast}$ from top to bottom) for the full sample, the subsamples with different stellar age (young, intermediate, old), the subsamples of different morphological types (ETGs and LTGs), the satellites, and the slow rotators. The classifications of subsamples are presented in \autoref{sec:sps} and \autoref{sec:morph_env_lambda}. For the FP, MP, total M/L, and total density slopes, the columns from left to right are: (1) the sample; (2) the function of the relations to be fitted with; (3) the best-fitting parameters; (4) the figures in which the relations are presented. For the dark matter fraction, we present the [10th, 16th, 50th, 84th, 90th] percentile values for both $\rm Qual\geqslant1$ and $\rm Qual=3$. The $L$, $M_{\rm JAM}$, $M_{\ast}$, and $(M/L)_{\rm JAM}$ are in solar units, while the units of velocity dispersion ($\sigma_{\rm e}$ and $\sigma_{\rm e}^{\rm intr}$) and size ($R_{\rm e}$ and $R_{\rm e}^{\rm maj}$) are ${\rm km\,s^{-1}}$ and $\rm kpc$ respectively. For the linear relations, $\Delta$ is the observed rms scatter derived from the \textsc{lts\_planefit} or \textsc{lts\_linefit} procedures.}
\label{tab:relations}
\setlength{\tabcolsep}{2.0mm}{
    \begin{tabular}{|c|l|l|c|}
    \hline \hline
    \multicolumn{4}{|c}{The Fundamental Plane and deprojected Fundamental Plane}\\
    \hline
    Sample & Function & Parameters & Ref\\
    \hline
    ETGs & $\lg\,L=a+b\times(\lg\,\sigma_{\rm e}-x_{0})+c\times(\lg\,R_{\rm e}-y_{0})$ &$a=10.5551$, $b=0.982$, $c=1.026$, $x_{0}=2.11$, $y_{0}=0.80$, $\Delta=0.13$&\autoref{fig:FP}\\
    LTGs & $\lg\,L=a+b\times(\lg\,\sigma_{\rm e}-x_{0})+c\times(\lg\,R_{\rm e}-y_{0})$ & $a=10.5994$, $b=1.590$, $c=1.068$, $x_{0}=2.11$, $y_{0}=0.80$, $\Delta=0.17$&\autoref{fig:FP}\\
    ETGs & $\lg\,L=a+b\times(\lg\,\sigma_{\rm e}^{\rm intr}-x_{0})+c\times(\lg\,R_{\rm e}^{\rm maj}-y_{0})$ &$a=10.4873$, $b=0.881$, $c=1.063$, $x_{0}=2.11$, $y_{0}=0.80$, $\Delta=0.14$&\autoref{fig:FP}\\
    LTGs & $\lg\,L=a+b\times(\lg\,\sigma_{\rm e}^{\rm intr}-x_{0})+c\times(\lg\,R_{\rm e}^{\rm maj}-y_{0})$ &$a=10.3738$, $b=1.986$, $c=0.635$, $x_{0}=2.11$, $y_{0}=0.80$, $\Delta=0.18$&\autoref{fig:FP}\\
    \hline \hline
    \multicolumn{4}{|c}{The Mass Plane and deprojected Mass Plane}\\
    \hline
    Sample & Function & Parameters & Ref\\
    \hline
    ETGs & $\lg\,M_{\rm JAM}=a+b\times(\lg\,\sigma_{\rm e}-x_{0})+c\times(\lg\,R_{\rm e}^{\rm maj}-y_{0})$ &$a=11.0432$, $b=1.985$, $c=0.9428$, $x_{0}=2.11$, $y_{0}=0.80$, $\Delta=0.067$&\autoref{fig:MP}\\
    LTGs & $\lg\,M_{\rm JAM}=a+b\times(\lg\,\sigma_{\rm e}-x_{0})+c\times(\lg\,R_{\rm e}^{\rm maj}-y_{0})$ &$a=11.0136$, $b=1.948$, $c=1.000$, $x_{0}=2.11$, $y_{0}=0.80$, $\Delta=0.11$&\autoref{fig:MP}\\
    ETGs & $\lg\,M_{\rm JAM}=a+b\times(\lg\,\sigma_{\rm e}^{\rm intr}-x_{0})+c\times(\lg\,R_{\rm e}^{\rm maj}-y_{0})$ &$a=10.9983$, $b=2.056$, $c=0.9221$, $x_{0}=2.11$, $y_{0}=0.80$, $\Delta=0.071$&\autoref{fig:MP}\\
    LTGs & $\lg\,M_{\rm JAM}=a+b\times(\lg\,\sigma_{\rm e}^{\rm intr}-x_{0})+c\times(\lg\,R_{\rm e}^{\rm maj}-y_{0})$ &$a=10.9592$, $b=2.080$, $c=0.8608$, $x_{0}=2.11$, $y_{0}=0.80$, $\Delta=0.068$&\autoref{fig:MP}\\
    \hline \hline
    \multicolumn{4}{|c}{The $\lg\,(M/L)_{\rm JAM}-\lg\,\sigma_{\rm e}$ relations}\\
    \hline
    Sample & Function & Parameters & Ref\\
    \hline
    Full & $\lg\,(M/L)_{\rm JAM}=\lg\,(M/L)_{0}+A\times{(\lg\,\sigma_{\rm e}-\lg\,\sigma_{0})}^{2}$ & $\lg\,(M/L)_{0}=0.51$, $A=1.03$, $\lg\,\sigma_{0}=1.84$ &\autoref{fig:MLjam_sigma}\\
    Old & $\lg\,(M/L)_{\rm JAM}=a+b\times(\lg\,\sigma_{\rm e}-x_{0})$ & $a=0.6329$, $b=0.655$, $x_{0}=2.11$, $\Delta=0.11$ &\autoref{fig:MLjam_sigma_agetype}\\
    Intermediate & $\lg\,(M/L)_{\rm JAM}=a+b\times(\lg\,\sigma_{\rm e}-x_{0})$ & $a=0.5822$, $b=0.417$, $x_{0}=2.11$, $\Delta=0.16$ &\autoref{fig:MLjam_sigma_agetype}\\
    Young & $\lg\,(M/L)_{\rm JAM}=a+b\times(\lg\,\sigma_{\rm e}-x_{0})$ & $a=0.5290$, $b=0.028$, $x_{0}=2.11$, $\Delta=0.20$ &\autoref{fig:MLjam_sigma_agetype}\\
    ETGs & $\lg\,(M/L)_{\rm JAM}=a+b\times(\lg\,\sigma_{\rm e}-x_{0})$ & $a=0.5739$, $b=0.893$, $x_{0}=2.11$, $\Delta=0.12$ &\autoref{fig:MLjam_sigma_type}\\
    LTGs & $\lg\,(M/L)_{\rm JAM}=\lg\,(M/L)_{0}+A\times{(\lg\,\sigma_{\rm e}-\lg\,\sigma_{0})}^{2}$ & $\lg\,(M/L)_{0}=0.55$, $A=1.00$, $\lg\,\sigma_{0}=1.89$ &\autoref{fig:MLjam_sigma_type}\\
    Slow rotators & $\lg\,(M/L)_{\rm JAM}=a+b\times(\lg\,\sigma_{\rm e}-x_{0})$ & $a=0.5925$, $b=0.877$, $x_{0}=2.11$, $\Delta=0.085$ &\autoref{fig:MLjam_sigma_type}\\
    Satellites& $\lg\,(M/L)_{\rm JAM}=\lg\,(M/L)_{0}+A\times{(\lg\,\sigma_{\rm e}-\lg\,\sigma_{0})}^{2}$ & $\lg\,(M/L)_{0}=0.52$, $A=0.85$, $\lg\,\sigma_{0}=1.78$ &\autoref{fig:MLjam_sigma_type}\\
    
    \hline \hline
    \multicolumn{4}{|c}{The $\overline{\gamma_{_{\rm T}}}-\sigma_{\rm e}$ relations}\\
    \hline
    Sample & Function & Parameters & Ref\\
    \hline
    Full & $\overline{\gamma_{_{\rm T}}} = A_{0}\left(\frac{\sigma_{\rm e}}{\sigma_{\rm b}}\right)^{\gamma}\left[\frac{1}{2}+\frac{1}{2}\left(\frac{\sigma_{\rm e}}{\sigma_{\rm b}}\right)^{\alpha}\right]^{\frac{\beta-\gamma}{\alpha}}$ & $A_{0}=2.18$, $\sigma_{\rm b}=189$, $\alpha=11.13$, $\beta=-0.02$,$\gamma=0.30$&\autoref{fig:gammat_sigma}\\
    Old &$\overline{\gamma_{_{\rm T}}} = A_{0}\left(\frac{\sigma_{\rm e}}{\sigma_{\rm b}}\right)^{\gamma}\left[\frac{1}{2}+\frac{1}{2}\left(\frac{\sigma_{\rm e}}{\sigma_{\rm b}}\right)^{\alpha}\right]^{\frac{\beta-\gamma}{\alpha}}$ & $A_{0}=2.20$, $\sigma_{\rm b}=179$, $\alpha=3.12$, $\beta=-0.10$, $\gamma=0.20$&\autoref{fig:gammat_sigma_agetype}\\
    Intermediate & $\overline{\gamma_{_{\rm T}}}=a+b\times(\lg\,\sigma_{\rm e}-x_{0})$ &$a=1.9711$, $b=0.596$, $x_{0}=2.11$, $\Delta=0.25$&\autoref{fig:gammat_sigma_agetype}\\
    Young & $\overline{\gamma_{_{\rm T}}}=a+b\times(\lg\,\sigma_{\rm e}-x_{0})$ &$a=1.8618$, $b=1.092$, $x_{0}=2.11$, $\Delta=0.27$&\autoref{fig:gammat_sigma_agetype}\\
    ETGs &$\overline{\gamma_{_{\rm T}}} = A_{0}\left(\frac{\sigma_{\rm e}}{\sigma_{\rm b}}\right)^{\gamma}\left[\frac{1}{2}+\frac{1}{2}\left(\frac{\sigma_{\rm e}}{\sigma_{\rm b}}\right)^{\alpha}\right]^{\frac{\beta-\gamma}{\alpha}}$ & $A_{0}=2.24$, $\sigma_{\rm b}=150$, $\alpha=397.85$, $\beta=-0.03$, $\gamma=0.11$&\autoref{fig:gammat_sigma_type}\\
    LTGs &$\overline{\gamma_{_{\rm T}}} = A_{0}\left(\frac{\sigma_{\rm e}}{\sigma_{\rm b}}\right)^{\gamma}\left[\frac{1}{2}+\frac{1}{2}\left(\frac{\sigma_{\rm e}}{\sigma_{\rm b}}\right)^{\alpha}\right]^{\frac{\beta-\gamma}{\alpha}}$ & $A_{0}=1.92$, $\sigma_{\rm b}=138$, $\alpha=14.27$, $\beta=0.14$, $\gamma=0.34$&\autoref{fig:gammat_sigma_type}\\
    Slow rotators &$\overline{\gamma_{_{\rm T}}} = A_{0}\left(\frac{\sigma_{\rm e}}{\sigma_{\rm b}}\right)^{\gamma}\left[\frac{1}{2}+\frac{1}{2}\left(\frac{\sigma_{\rm e}}{\sigma_{\rm b}}\right)^{\alpha}\right]^{\frac{\beta-\gamma}{\alpha}}$ & $A_{0}=2.22$, $\sigma_{\rm b}=174$, $\alpha=2.47$, $\beta=-0.20$, $\gamma=0.14$&\autoref{fig:gammat_sigma_type}\\
    Satellites &$\overline{\gamma_{_{\rm T}}} = A_{0}\left(\frac{\sigma_{\rm e}}{\sigma_{\rm b}}\right)^{\gamma}\left[\frac{1}{2}+\frac{1}{2}\left(\frac{\sigma_{\rm e}}{\sigma_{\rm b}}\right)^{\alpha}\right]^{\frac{\beta-\gamma}{\alpha}}$ & $A_{0}=2.19$, $\sigma_{\rm b}=175$, $\alpha=9.93$, $\beta=0.02$, $\gamma=0.29$&\autoref{fig:gammat_sigma_type}\\
    \hline \hline
    
    \multicolumn{4}{|c}{The $f_{\rm DM}(<R_{\rm e})-\lg\,M_{\ast}$ relations}\\
    \hline
    Sample & Percentile & Values (per cent) & Ref\\
    \hline
    Full & [10th, 16th, 50th, 84th, 90th] & $\rm Qual\geqslant1$: [0, 0, 7.6, 26, 37]; \, $\rm Qual=3$: [0, 0, 5.2, 17, 22] &\autoref{fig:fdm_Ms_qual}\\
    Old & [10th, 16th, 50th, 84th, 90th] & $\rm Qual\geqslant1$: [0, 0, 7.3, 20, 25]; \, $\rm Qual=3$: [0, 0, 5.5, 14, 19] &\autoref{fig:fdm_Ms_agetype}\\
    Intermediate & [10th, 16th, 50th, 84th, 90th] & $\rm Qual\geqslant1$: [0, 0, 7.7, 29, 38]; \, $\rm Qual=3$: [0, 0, 5.9, 17, 22] &\autoref{fig:fdm_Ms_agetype}\\
    Young & [10th, 16th, 50th, 84th, 90th] & $\rm Qual\geqslant1$: [0, 0, 8.4, 53, 77]; \, $\rm Qual=3$: [0, 0, 2.8, 20, 28] &\autoref{fig:fdm_Ms_agetype}\\
    ETGs & [10th, 16th, 50th, 84th, 90th] & $\rm Qual\geqslant1$: [0, 0, 6.5, 18, 23]; \, $\rm Qual=3$: [0, 0, 4.6, 12, 14] &\autoref{fig:fdm_Ms_type}\\
    LTGs & [10th, 16th, 50th, 84th, 90th] & $\rm Qual\geqslant1$: [0, 0, 9.4, 43, 63]; \, $\rm Qual=3$: [0, 0, 4.7, 19, 25] &\autoref{fig:fdm_Ms_type}\\
    Slow rotators & [10th, 16th, 50th, 84th, 90th] & $\rm Qual\geqslant1$: [0, 0, 7.9, 21, 30]; \, $\rm Qual=3$: None &\autoref{fig:fdm_Ms_type}\\
    Satellites & [10th, 16th, 50th, 84th, 90th] & $\rm Qual\geqslant1$: [0, 0, 7.4, 29, 38]; \, $\rm Qual=3$: [0, 0, 5.3, 17, 22] &\autoref{fig:fdm_Ms_type}\\
    \hline \hline
    \end{tabular}}
\end{table*}

We summarize the main results as below:
\begin{itemize}[align=left,leftmargin=2em,itemsep=1em]
    \item We confirm that the deprojected MPs for both ETGs and LTGs, which have been corrected for the inclination effect, agree very well with the virial predictions in terms of the coefficients ($b\approx2$, $c\approx1$) and the negligible intrinsic scatter (middle panels in \autoref{fig:MP}). This confirms previous findings that the tilt and the scatter of the FP are mainly due to the variation of total $\rm M/L$ along and perpendicular to the FP, while the effect of non-homology in light profiles (captured by the Sersic index) is negligible (bottom panels in \autoref{fig:MP}). The variation of total M/L for ETGs is dominated by the stellar mass-to-light ratio $M_{\ast}/L$ variation (captured by the stellar age), while the one for LTGs can be attributed to the $M_{\ast}/L$ variation at $L>10^{10.2}{\rm L_{\odot,r}}$ and the $f_{\rm DM}(<R_{\rm e})$ variation at $L<10^{10.2}{\rm L_{\odot,r}}$ (\autoref{fig:FP}).  
    
    \item We measure a clear parabolic variation in the total mass-to-light ratios $\rm M/L$ variation with $\sigma_{\rm e}$: the total M/L is larger for the galaxies with higher $\sigma_{\rm e}$ (see \autoref{fig:MLjam_sigma} and \autoref{eq:MLdyn_fit}). For the galaxies with different stellar ages, the $M/L-\sigma_{\rm e}$ relations can be described as straight lines with different slopes and the slopes become steeper for the older galaxies (\autoref{fig:MLjam_sigma_agetype}). The ETGs and slow rotators have nearly linear $M/L-\sigma_{\rm e}$ relations, while the relations of LTGs and the satellites are similar to the one for the full sample (\autoref{fig:MLjam_sigma_type}). 
    
    \item We confirm and improve previous determinations of the relation between the mass-weighted total density slopes $\overline{\gamma_{_{\rm T}}}$ and $\sigma_{\rm e}$. Our best fitting relation has the form of \autoref{eq:gammat_fit} (see \autoref{fig:gammat_sigma}): the $\overline{\gamma_{_{\rm T}}}$ gets steeper with increasing $\sigma_{\rm e}$ until $\lg\,(\sigma_{\rm e}/{\rm km\,s^{-1}})\approx2.25$, above which the $\overline{\gamma_{_{\rm T}}}$ remain unchanged with good accuracy at the "universal" value $\overline{\gamma_{_{\rm T}}}\approx2.2$ reported by previous studies. We additionally look for trends as a function of stellar age and find that the trend varies with the mean age of the stellar population. At fixed $\sigma_{\rm e}$, the $\overline{\gamma_{_{\rm T}}}$ is steeper for the older population. The slopes of $\overline{\gamma_{_{\rm T}}}-\sigma_{\rm e}$ relations become shallower with increasing stellar age, while the turnover of the $\overline{\gamma_{_{\rm T}}}-\sigma_{\rm e}$ relation only exists for the old galaxies (\autoref{fig:gammat_sigma_agetype}). We also find that the LTGs have systematically shallower total slopes than the ETGs and the satellites have systematically steeper ($\approx0.1$) than the full sample (dominated by central galaxies). 
    
    \item We show the dark matter fraction relations using two mass models and confirm that our $f_{\rm DM}(<R_{\rm e})-M_{\ast}$ relations are not affected by the model differences (\autoref{fig:fdm_Ms_qual}). The $f_{\rm DM}(<R_{\rm e})$ decreases with increasing $M_{\ast}$ until $M_{\ast}=10^{10}{\rm M_{\odot}}$, above which the $f_{\rm DM}(<R_{\rm e})$ remains unchanged and small ($\approx10$ per cent). However, we highlight for the first time that $\sigma_{\rm e}$ or the age of the stellar population are better predictors of $f_{\rm DM}(<R_{\rm e})$ than the stellar mass that is generally used. The dark matter fractions increase to a median of $f_{\rm DM}(<R_{\rm e})=33$ percent for galaxies with $\sigma_{\rm e}\la100$ \kms. We find that only young galaxies show a strong dependence of $f_{\rm DM}(<R_{\rm e})$ on the $M_{\ast}$, while the intermediate and old galaxies have invariant low dark matter fraction (\autoref{fig:fdm_Ms_agetype}). A significant difference in the relations between ETGs and LTGs is observed: the ETGs have invariant low dark matter fractions (a median of 7 per cent), while the LTGs show a decreasing trend with increasing $M_{\ast}$ (\autoref{fig:fdm_Ms_type}). The above results do not change when only using the best quality ($\rm Qual=3$) sample (the black solid curves in \autoref{fig:fdm_Ms_qual}, \autoref{fig:fdm_Ms_agetype} and \autoref{fig:fdm_Ms_type}), although the $\rm Qual=3$ sample only covers a stellar mass range of $M_{\ast}=10^{10-11.5}{\rm M_{\odot}}$. 
    
    \item We incorporate the stellar mass-to-light ratio gradients (taken from the stellar population analysis in \citetalias{Lu2023a}) into the dynamical models to test the effect of spatially constant $M_{\ast}/L$ assumption (\autoref{sec:fdm_MLgrad}). If we assume that the galaxies have the same $M_{\ast}/L$ gradients as inferred from the SPS models, the $f_{\rm DM}(<R_{\rm e})$ increase by $\sim 7$ per cent for the NFW models ($\sim 13$ per cent for the gNFW models). The trend of $f_{\rm DM}(<R_{\rm e})-M_{\ast}$ relation does not change qualitatively under this assumption of $M_{\ast}/L$ gradients (\autoref{fig:fdm_Ms_MLgrad}). 
    
    \item The dynamical properties ($\sigma_{\rm e}$, $\lambda_{\rm R_{e}}$, $\overline{\gamma_{_{\rm T}}}$, and $f_{\rm DM}(<R_{\rm e})$) on the ($M_{\rm JAM}-R_{\rm e}^{\rm maj}$) plane (\autoref{fig:mass_size_plane}) can be qualitatively interpreted by the scenario of two evolutionary channels: (i) the bulge growth (through gas accretion or gas-rich mergers) moving the galaxies from left to right, while increasing the $\sigma_{\rm e}$, making the $\overline{\gamma_{_{\rm T}}}$ steeper, reducing the central dark matter fraction, leaving the $\lambda_{\rm R_{e}}$ nearly unchanged; (ii) the dry mergers moving the galaxies along the constant $\sigma_{\rm e}$ lines upwards, while decreasing the $\lambda_{\rm R_{e}}$, changing the $\overline{\gamma_{_{\rm T}}}$ to be nearly isothermal, and leaving the dark matter fractions unchanged.
\end{itemize}

\section*{Acknowledgements}
We acknowledge the support of National Nature Science Foundation of China (Nos 11988101,12022306), the National Key R$\&$D Program of China No. 2022YFF0503403,  the support from the Ministry of Science and Technology of China (Nos. 2020SKA0110100),  the science research grants from the China Manned Space Project (Nos. CMS-CSST-2021-B01,CMS-CSST-2021-A01), CAS Project for Young Scientists in Basic Research (No. YSBR-062), and the support from K.C.Wong Education Foundation. SM acknowledges the National Key Research and Development Program of China (No. 2018YFA0404501 to SM), the National Science Foundation of China (Grant No. 11821303, 11761131004 and 11761141012), the Tsinghua University Initiative Scientific Research Program ID 2019Z07L02017, and the science research grants from the China Manned Space Project with NO. CMS-CSST-2021-A11.

Funding for the Sloan Digital Sky 
Survey IV has been provided by the 
Alfred P. Sloan Foundation, the U.S. 
Department of Energy Office of 
Science, and the Participating 
Institutions. 

SDSS-IV acknowledges support and 
resources from the Center for High 
Performance Computing  at the 
University of Utah. The SDSS 
website is www.sdss.org.

SDSS-IV is managed by the 
Astrophysical Research Consortium 
for the Participating Institutions 
of the SDSS Collaboration including 
the Brazilian Participation Group, 
the Carnegie Institution for Science, 
Carnegie Mellon University, Center for 
Astrophysics | Harvard \& 
Smithsonian, the Chilean Participation 
Group, the French Participation Group, 
Instituto de Astrof\'isica de 
Canarias, The Johns Hopkins 
University, Kavli Institute for the 
Physics and Mathematics of the 
Universe (IPMU) / University of 
Tokyo, the Korean Participation Group, 
Lawrence Berkeley National Laboratory, 
Leibniz Institut f\"ur Astrophysik 
Potsdam (AIP),  Max-Planck-Institut 
f\"ur Astronomie (MPIA Heidelberg), 
Max-Planck-Institut f\"ur 
Astrophysik (MPA Garching), 
Max-Planck-Institut f\"ur 
Extraterrestrische Physik (MPE), 
National Astronomical Observatories of 
China, New Mexico State University, 
New York University, University of 
Notre Dame, Observat\'ario 
Nacional / MCTI, The Ohio State 
University, Pennsylvania State 
University, Shanghai 
Astronomical Observatory, United 
Kingdom Participation Group, 
Universidad Nacional Aut\'onoma 
de M\'exico, University of Arizona, 
University of Colorado Boulder, 
University of Oxford, University of 
Portsmouth, University of Utah, 
University of Virginia, University 
of Washington, University of 
Wisconsin, Vanderbilt University, 
and Yale University.

\section*{Data Availability}
The catalogues of dynamical quantities (\citetalias{Zhu2023a}) and the stellar population properties (\citetalias{Lu2023a}) are publicly available on the website of MaNGA DynPop (\url{https://manga-dynpop.github.io}). The catalogue of dynamical properties is also publicly available on the journal website, as a supplementary file of \citetalias{Zhu2023a}.

\section*{Software Citations}
This work uses the following software packages:

\begin{itemize}

\item
\href{https://github.com/astropy/astropy}{{Astropy}}
\citep{astropy1, astropy2}

\item
\href{https://github.com/matplotlib/matplotlib}{{Matplotlib}}
\citep{Matplotlib2007}

\item
\href{https://github.com/numpy/numpy}{{NumPy}}
\citep{Numpy2011}

\item
\href{https://www.python.org/}{{Python}}
\citep{Python3}

\item
\href{https://github.com/scikit-image/scikit-image}{{Scikit-image}}
\citep{scikit-image}

\item
\href{https://github.com/scipy/scipy}{{Scipy}}
\citep{Scipy2020}

\item
\href{https://pypi.org/project/ltsfit/}{{LtsFit}}
\citep{Cappellari2013a}

\item
\href{https://pypi.org/project/loess/}{{LOESS}}
\citep{Cappellari2013b}

\end{itemize}


\bibliographystyle{mnras}
\bibliography{ref} 



\appendix
\section{Effects of the \texorpdfstring{$\rm Qual=0$}{} galaxies on the FP, MP, and \texorpdfstring{$M/L$}{}}
\label{appendix:Qual0}
\begin{figure*}
    \centering
    \includegraphics[width=0.85\textwidth]{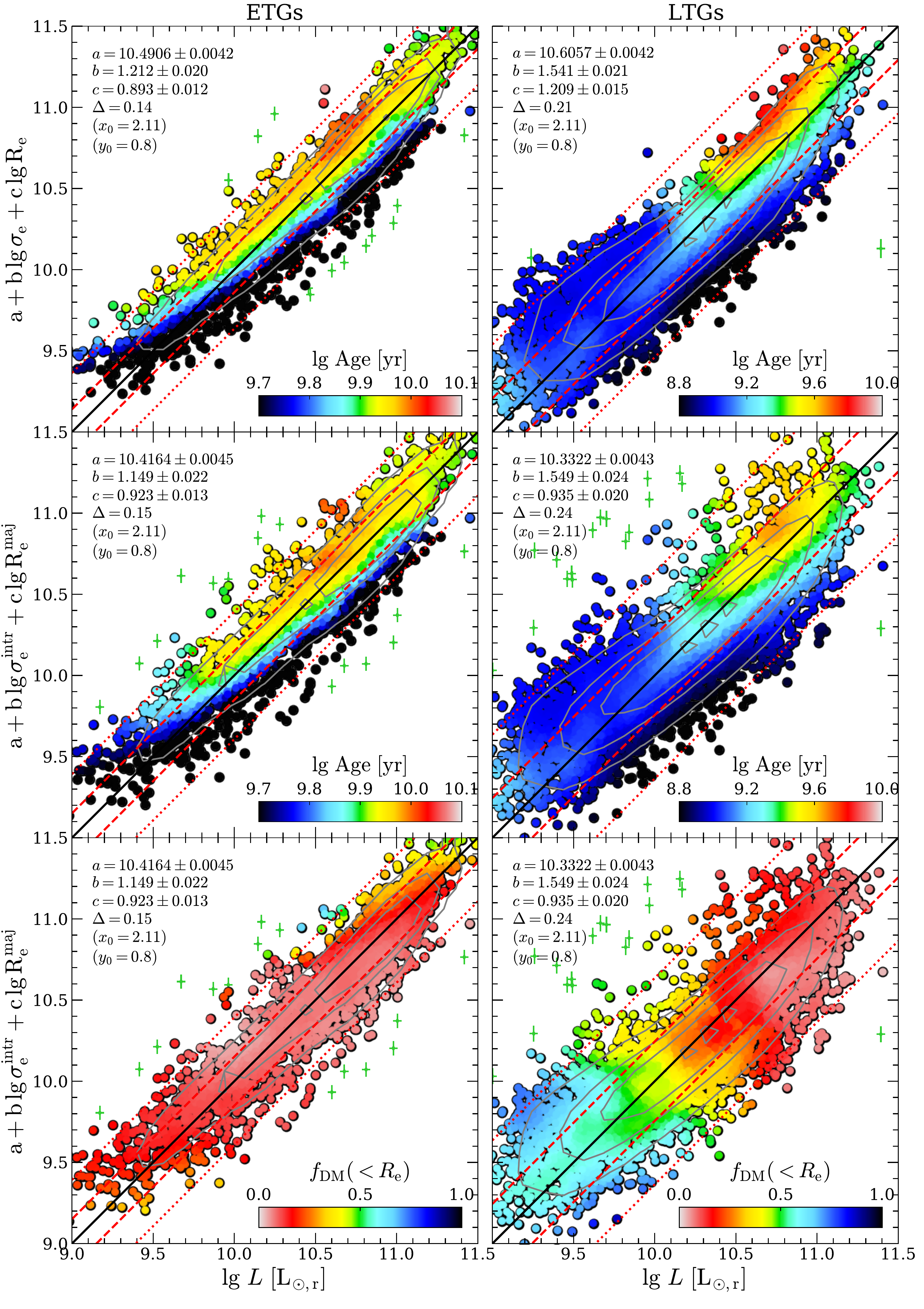}
    \caption{The same as \autoref{fig:FP}, but for the $\rm Qual \geqslant0$ sample.}
    \label{fig:FP_Qual0}
\end{figure*}

\begin{figure*}
    \centering
    \includegraphics[width=0.85\textwidth]{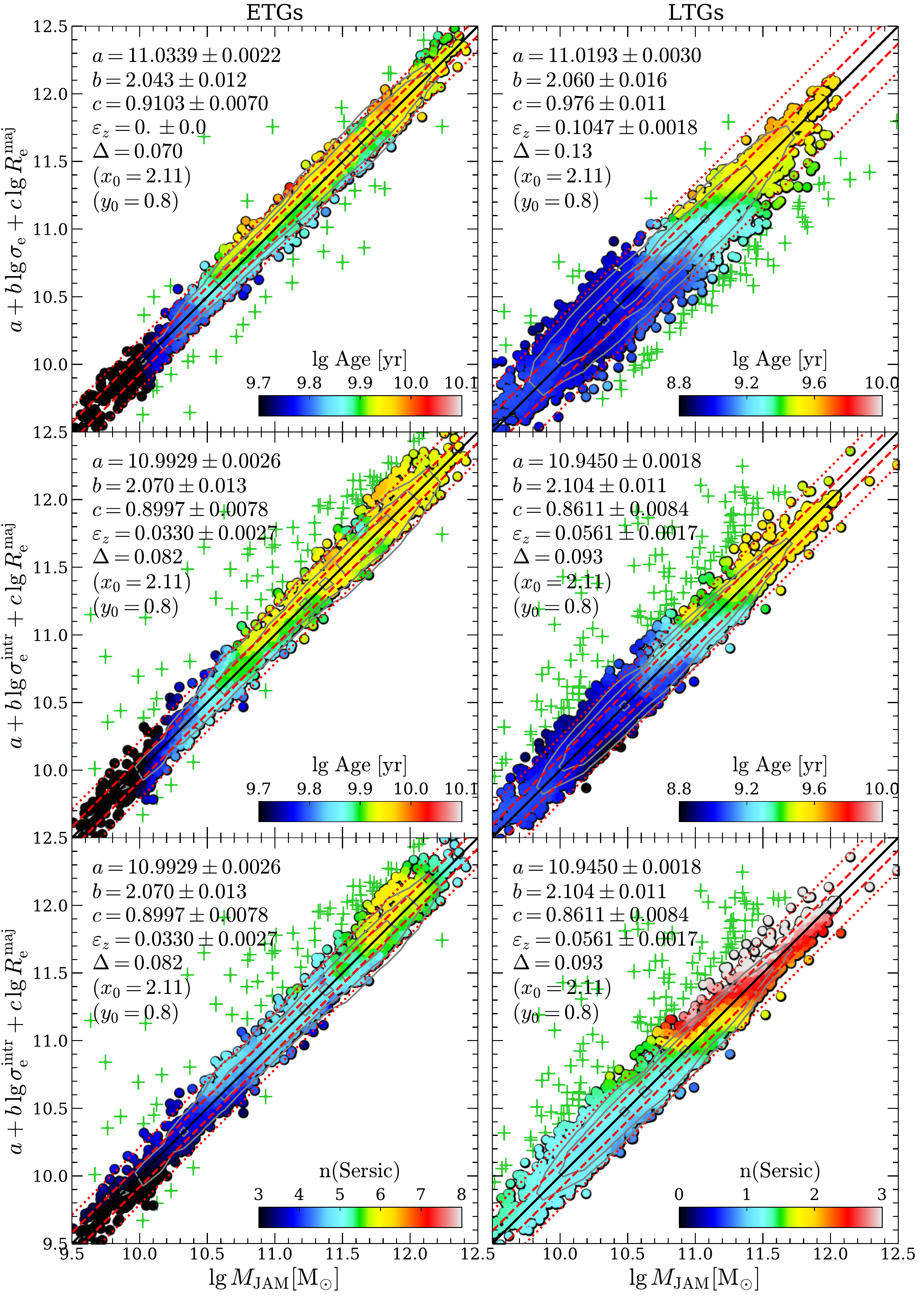}
    \caption{The same as \autoref{fig:MP}, but for the $\rm Qual \geqslant0$ sample.}
    \label{fig:MP_Qual0}
\end{figure*}

\begin{figure}
    \centering
    \includegraphics[width=\columnwidth]{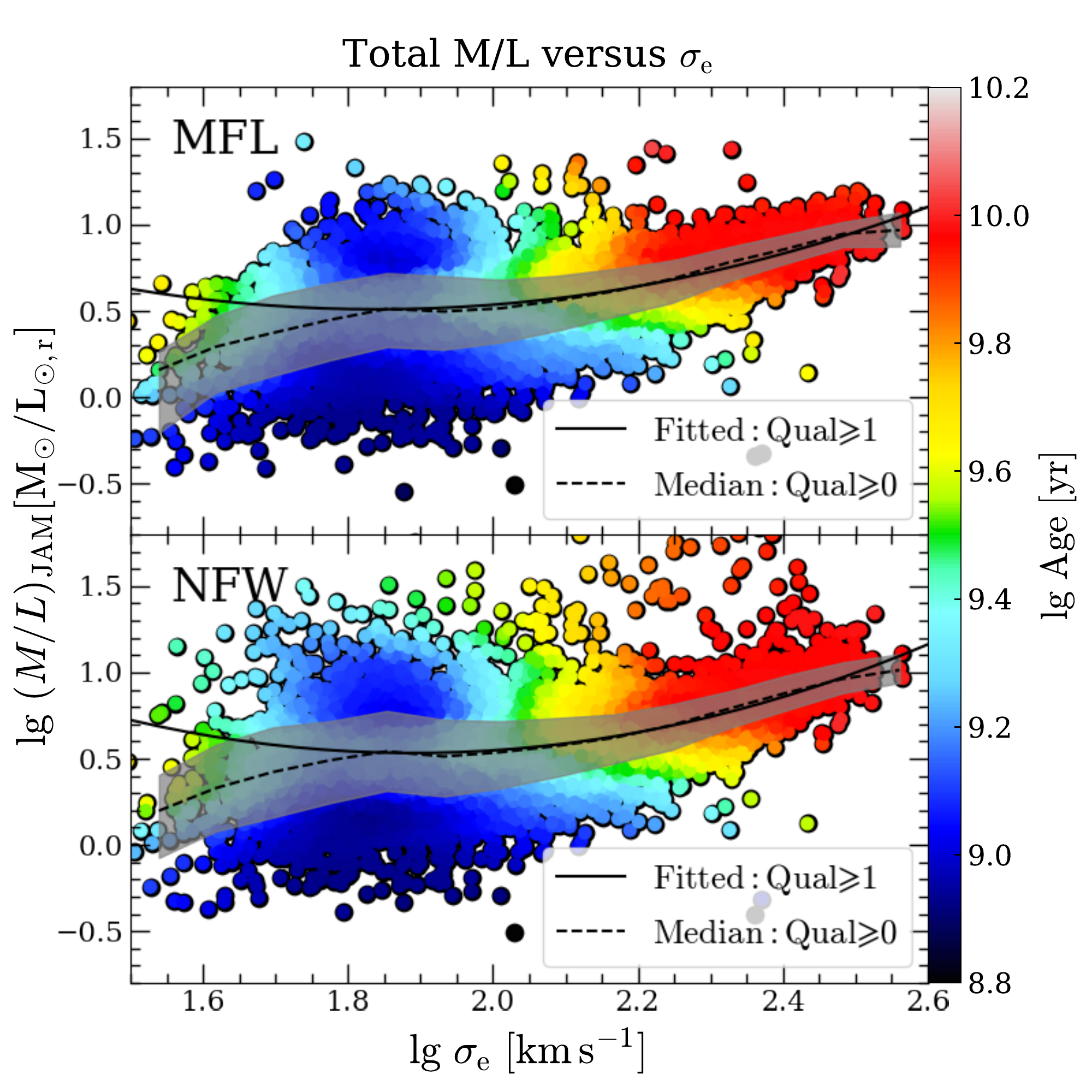}
    \caption{The same as \autoref{fig:MLjam_sigma}, but for the $\rm Qual\geqslant0$ sample. The best-fitting relation for the $\rm Qual\geqslant1$ galaxies (black solid line) is consistent with the median relation of the $\rm Qual\geqslant0$ sample (black dashed line) at $\sigma_{\rm e} \gtrsim 60\,{\rm km\,s^{-1}}$.}
    \label{fig:MLjam_sigma_Qual0}
\end{figure}

We present the FP, MP, and $(M/L)_{\rm e}-\sigma_{\rm e}$ relation for the $\rm Qual\geqslant0$ sample (9360 galaxies) in \autoref{fig:FP_Qual0}, \autoref{fig:MP_Qual0}, and \autoref{fig:MLjam_sigma_Qual0}, respectively. With the $\rm Qual=0$ sample included, the FP and MP for the ETGs remain nearly unchanged, while the planes for the LTGs have a larger scatter. In this case, the distributions of stellar age, Sersic index, and dark matter fraction on the FP and MP are similar, suggesting that the conclusions in \autoref{sec:FP_MP} still hold when including the $\rm Qual=0$ sample. Moreover, we find that the $(M/L)_{\rm e}-\sigma_{\rm e}$ relation for the $\rm Qual\geqslant0$ sample is consistent with the one for the $\rm Qual\geqslant1$ sample (\autoref{sec:ML}) at $\sigma_{\rm e} \gtrsim 60\,{\rm km\,s^{-1}}$, below which the $\rm Qual=0$ galaxies with low-$\sigma_{\rm e}$ systematically have smaller $M/L$ than $\rm Qual\geqslant1$ galaxies.


\section{The original scatter orthogonal to the {\sc loess}-smoothed scaling relations}
\label{appendix:scatter_loess}
A two-dimensional {\sc loess}-smoothed map is a way of visualizing the average value of a function that depends on two variables. It is similar to the average trend that is often shown in one-dimensional plots, but in two dimensions. {\sc loess} is a more robust and accurate method than a simple average because it takes into account the variation around each point. However, unlike a one-dimensional plot, it is difficult to visualize both the mean value and the scatter in a two-dimensional map. One way to show the scatter would be to use non-smoothed maps with colours, but this can be confusing for the human brain. A better way is to show an approximate edge-on view of the average trends computed by {\sc loess}.

In general, the average trends in two dimensions do not need to follow simple planes but will be described by more complex surfaces. This implies that there may not be a single direction that shows the surfaces edge-on. However, we can show the surfaces along an approximate direction that minimizes the scatter around the {\sc loess} surface. To achieve this, we use \textsc{lts\_planefit} software to fit a plane to the {\sc loess}-smoothed values $Z_{\rm loess} = a+b\times (X-X_0)+c\times (Y-Y_0)$ and then rotate best-fitting plane to be edge-on view.

\begin{figure*}
    \centering
    \includegraphics[width=2\columnwidth]{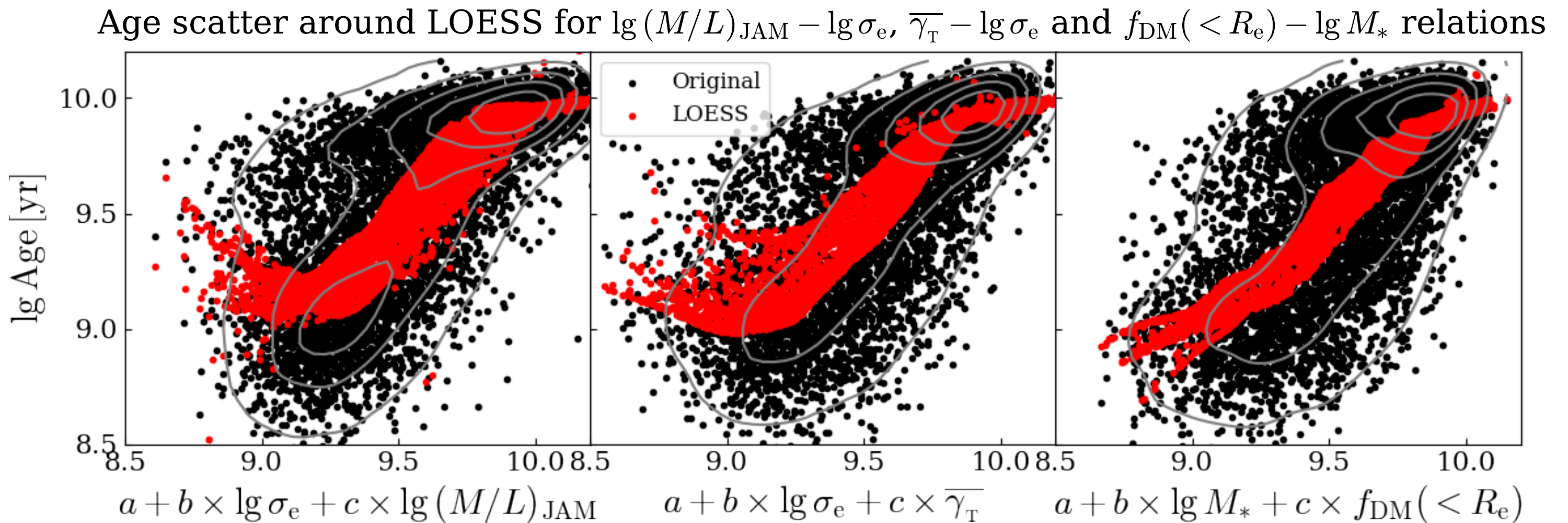}
    \caption{From left to right, the panels show the stellar age scatter orthogonal to the $\lg\,(M/L)_{\rm JAM}-\lg\,\sigma_{\rm e}$, $\overline{\gamma_{_{\rm T}}}-\lg\,\sigma_{\rm e}$ and $f_{\rm DM}(<R_{\rm e})-M_{\ast}$ relations, corresponding to edge-on view of the top panel of \autoref{fig:MLjam_sigma}, the bottom panel of \autoref{fig:gammat_sigma}, and the top right panel of \autoref{fig:fdm_Ms_qual}. The angle of edge-on view is determined from fitting a plane $Z_{\rm loess} = a+b\times (X-X_0)+c\times (Y-Y_0)$ to the \textsc{loess}-smoothed values ($X$, $Y$, $Z_{\rm loess}$) (see the text in \aref{appendix:scatter_loess}). The black symbols denote the original data (the grey contours show the kernel density estimate), while the red symbols correspond to the \textsc{loess}-smoothed data.}
    \label{fig:JAMprop_edgeon}
\end{figure*}

\begin{figure*}
    \centering
    \includegraphics[width=1.6\columnwidth]{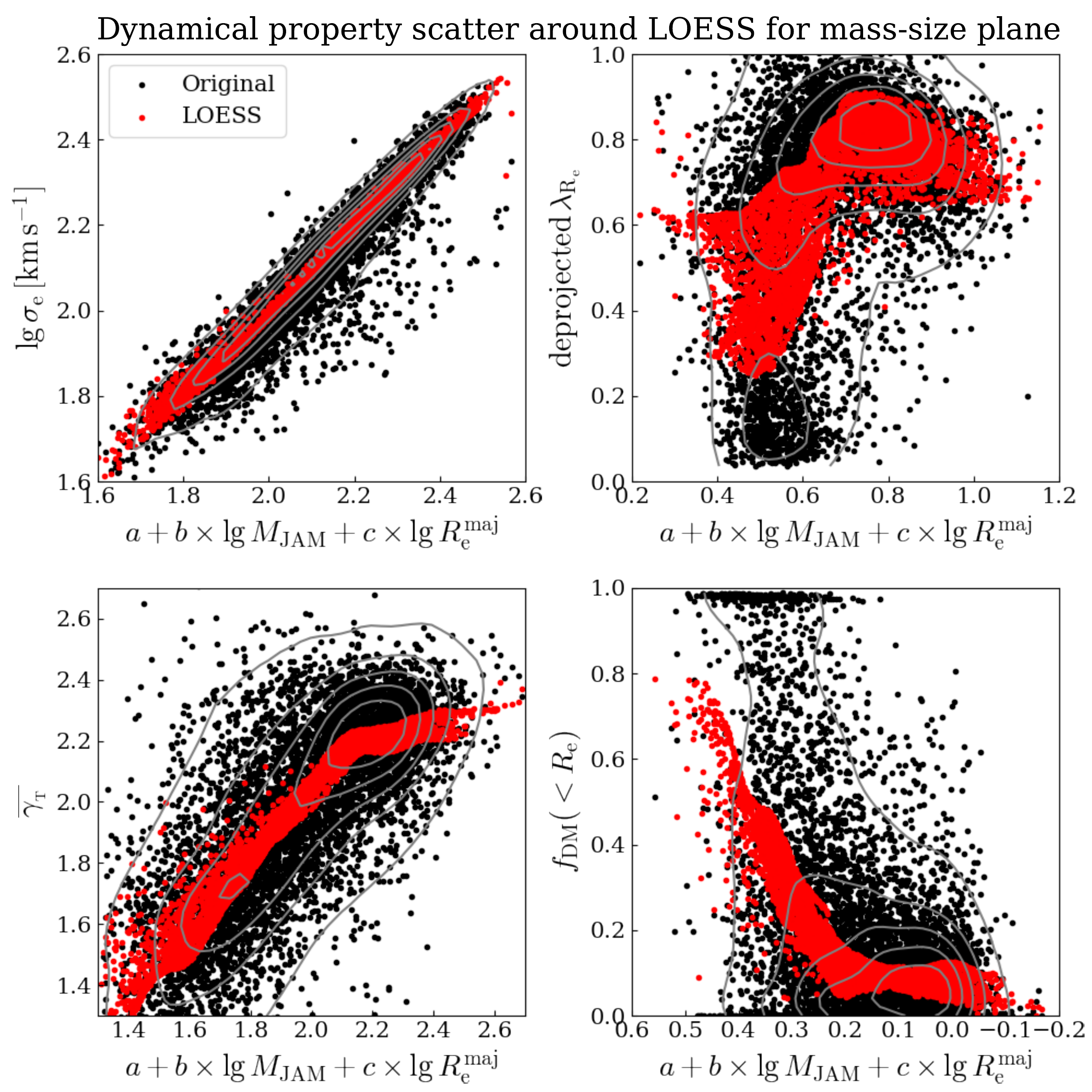}
    \caption{The scatter of galaxy properties on the mass-size plane. The panels are similar to \autoref{fig:mass_size_plane}, but in the edge-on view of ($M_{\rm JAM}$, $R_{\rm e}^{\rm maj}$, $Z_{\rm loess}$) planes, where $Z_{\rm loess}$ are the \textsc{loess}-smoothed $\sigma_{\rm e}$, deprojected $\lambda_{\rm R_{e}}$, $\overline{\gamma_{_{\rm T}}}$, and $f_{\rm DM}(<R_{\rm e})$. The symbols and contours are similar to \autoref{fig:JAMprop_edgeon}.}
    \label{fig:Mass_size_plane_edgeon}
\end{figure*}

The results are shown in \autoref{fig:JAMprop_edgeon} and \autoref{fig:Mass_size_plane_edgeon} for all plots where we show {\sc loess}-smoothed quantities in the main text. These plots allow one to visually assess the scatter around the best-fitting {\sc loess} trends. The {\sc loess}-smoothed trends follow the original trends very well in all plots except for the bottom right panel of \autoref{fig:Mass_size_plane_edgeon}, which seemingly shows a slight offset. However, as clearly showed in fig.~2 of \citet{Lu2023b}, the red points in this plot approximately follow the original $f_{\rm DM}(<R_{\rm e})-\lg\,\sigma_{\rm e}$ relation (we confirmed that there is a strong linear anti-correlation between $\lg\,\sigma_{\rm e}$ and the best-fitting $a+b\times\lg\,M_{\rm JAM}+c\times\lg\,R_{\rm e}^{\rm maj}$), while the cluster of black points at $f_{\rm DM}(<R_{\rm e})=1$ are outliers which can be clipped with the $\lg\,(M_{\ast}/L)-\lg\,\sigma_{\rm e}$ relation \citep[fig.~1]{Lu2023b}.


\bsp	
\label{lastpage}
\end{document}